\documentclass[iop,11pt,twocolumn]{emulateapj}
\usepackage{natbib}
\usepackage{color}
\newcommand{\HI}{H\,{\scriptsize I}}

\newcommand{\MgII}{Mg${{\rm I\hspace{-.1em}I}}$ }

\newcommand{\SFD}{\mathrm{SFD}}
\newcommand{\Ebv}{E(B-V)}

\newcommand{\mum}{\ensuremath{\mu \mathrm{m}}}

\slugcomment{}
\shortauthors{Kashiwagi et al.}
\shorttitle{Modeling anomaly in the Galactic extinction map}
\begin{document}
%%%%%%%%%%%%%%%%%%%%%%%%%%%%%%%%%%%%%%%%%%%%

%%%%%%%%%%%%%%%%%%%%%%%%%%%%%%%%%%%%%%%%%%%%
%      Title
%%%%%%%%%%%%%%%%%%%%%%%%%%%%%%%%%%%%%%%%%%%%
\title{
Modeling the anomaly of surface number densities of galaxies on the
Galactic extinction map due to their FIR emission contamination
}
%%%%%%%%%%%%%%%%%%%%%%%%%%%%%%%%%%%%%%%%%%%%
%      Author & Affiliation
%%%%%%%%%%%%%%%%%%%%%%%%%%%%%%%%%%%%%%%%%%%%
\author{
  Toshiya Kashiwagi\altaffilmark{1},
  Yasushi Suto\altaffilmark{1,2,3},
  Atsushi Taruya\altaffilmark{1,2,4,5},\\
  Issha Kayo\altaffilmark{6,7}, 
  Takahiro Nishimichi\altaffilmark{4,8}, and
  Kazuhiro Yahata\altaffilmark{1,9}
}
\altaffiltext{1}{Department of Physics, The University of Tokyo,
  Tokyo 113-0033, Japan}
\altaffiltext{2}{Research Center for the Early Universe, 
 School of Science, The University of Tokyo, Tokyo 113-0033, Japan}
\altaffiltext{3}{Department of Astrophysical Sciences, Princeton
  University, Princeton, NJ 08544}
\altaffiltext{4}{Institute for the Physics and Mathematics of the Universe,
 University of Tokyo, Kashiwa, Chiba 277-8568, Japan}
\altaffiltext{5}{present address: Yukawa Institute for Theoretical
  Physics, Kyoto University, Kyoto 606-8502, Japan}
\altaffiltext{6}{Department of Physics, Toho University,  Funabashi,
  Chiba 274-8510, Japan}
\altaffiltext{7}{Department of Liberal Arts, Tokyo University of Technology, Ota-ku, Tokyo 144-8650, Japan}
\altaffiltext{8}{present address: 
Institut d'Astrophysique de Paris, 98 bis Bd Arago, 75014 Paris, France}
\altaffiltext{9}{present address: Canon Inc. Ohta-ku, Tokyo 146-8501, Japan}
\email{kashiwagi@utap.phys.s.u-tokyo.ac.jp}

%%%%%%%%%%%%%%%%%%%%%%%%%%%%%%%%%%%%%%%%%%%%
%     Abstract & Keywords
%%%%%%%%%%%%%%%%%%%%%%%%%%%%%%%%%%%%%%%%%%%%
\begin{abstract}
The most widely used Galactic extinction map
\citep[][SFD]{Schlegel;1998} is constructed assuming that the observed
FIR fluxes entirely come from the Galactic dust.  According to the
earlier suggestion by \citet{Yahata;2007}, we consider how far-infrared
(FIR) emission of galaxies affects the SFD map.  We first compute the
surface number density of SDSS DR7 galaxies as a function of the
$r$-band extinction, $A_{r,\rm SFD}$.  We confirm that the surface
densities of those galaxies {\em positively} correlate with $A_{r,\rm
SFD}$ for $A_{r,\rm SFD}<0.1$, as first discovered by
\citet{Yahata;2007} for SDSS DR4 galaxies.  Next we construct an
analytic model to compute the surface density of galaxies taking account
of the contamination of their FIR emission. We adopt a log-normal
probability distribution for the ratio of $100\mum$ and $r$-band
luminosities of each galaxy, $y \equiv (\nu L)_{100\mum}/(\nu
L)_r$. Then we search for the mean and {\it r.m.s} values of $y$ that
fit the observed anomaly using the analytic model.  The required values
to reproduce the anomaly are roughly consistent with those measured from
the stacking analysis of SDSS galaxies \citep{Kashiwagi;2013}.  Due to
the limitation of our statistical modeling, we are not yet able to
remove the FIR contamination of galaxies from the extinction
map. Nevertheless the agreement with the model prediction suggests that
the FIR emission of galaxies is mainly responsible for the observed 
anomaly. While the corresponding systematic error in the Galactic
extinction map is 0.1 to 1mmag, it is directly correlated with galaxy
clustering, and thus needs to be carefully examined in precision
cosmology.
\end{abstract}
\keywords{dust, extinction --- large-scale structure of universe
 --- cosmology: observations}
%%%%%%%%%%%%%%%%%%%%%%%%%%%%%%%%%%%%%%%%%%%%%%

%%%%%%%%%%%%%%%%%%%%%%%%%%%%%%%%%%%%%%%%%%%%%%
\section{Introduction} 
\label{sec:intro}

The Galactic extinction map is the most fundamental data for astronomy
and cosmology, since all extragalactic astronomical observations are
inevitably conducted through the Galactic foreground, thus affected by
the Galactic interstellar dust.  In particular, lights in optical and
ultraviolet bands are dimmed by the absorption and scattering of the
Galactic dust.  Therefore, we cannot determine any fundamental
quantities such as intrinsic luminosities or colors of extragalactic
objects without proper correction for the dust extinction.  This is why
the Galactic extinction correction could be one of the most critical
sources of systematics.
 
The most widely-used Galactic extinction map was constructed by
\citet[][hereafter SFD]{Schlegel;1998} based on the IRAS/ISSA and
COBE/DIRBE Far-infrared (FIR) emission maps, which are dominated by
thermal dust emission.  The construction of the SFD map consists of the
 following procedures: 

\begin{enumerate}
\renewcommand{\labelenumi}{(\roman{enumi})}
\item constructing a dust temperature map from
the ratio of the 100 {\mum} flux to the 240 {\mum} flux measured by
DIRBE, which has $1^{\circ}.1$ FWHM spatial resolution, 
\item calibrating the ISSA 100 {\mum} emission map, which has the 
resolution of $6^{\prime}.1$ FWHM, according to the DIRBE 100 {\mum} map,
\item correcting the calibrated ISSA 100 {\mum} map for dust temperature 
using the previous temperature map, 
\item converting the ISSA 100 {\mum} 
map to color excess, $\Ebv$, assuming the proportionality between
the temperature corrected 100 {\mum} flux, $I_{100\mum}$, and the dust
column density:
%%%%%%%%%%%%%%%%%%%%%%%%%%%%%%%%%%%%%%%
\begin{equation}
\label{eq:SFD-extinction}
E(B-V) = pI_{100\mum} X(T),
\end{equation}
%%%%%%%%%%%%%%%%%%%%%%%%%%%%%%%%%%%%%%%
where $p$ is a constant determined from {\MgII}indices of
elliptical galaxies as standard color indicators, and $X(T)$ is the
correction for the dust temperature.
\end{enumerate}

The SFD map has achieved significant improvement in precision and resolution
compared to the previous extinction maps constructed from {\HI}
21-${\mathrm{cm}}$ emission \citep{Burstein;1978,Burstein;1982}.
Nevertheless, it should be noted that the map is not based on any
direct measurement of the dust {\it absorption}, but derived from its
{\it emission}. Indeed one needs several assumptions to convert the FIR
emission map into the extinction map.  This is why it is important to
test the reliability of the SFD map by comparing with other independent
observations.

In high-extinction regions, such as molecular clouds or near the
Galactic plane, many earlier studies examined the SFD map using star
counts, NIR galaxy colors, and galaxy number counts
\citep{Arce;1999a,Arce;1999b,Chen;1999,Cambresy;2001,Cambresy;2005,
Dobashi;2005, Yasuda;2007,Rowles;2009}.  They often report that the SFD
map over-predicts extinction in the high-extinction regions, possibly
because of the poor angular resolution of the dust temperature map
\citep{Arce;1999a,Arce;1999b} and/or the existence of cold dust
components with high emissivity in FIR.

In contrast, its reliability in low-extinction regions has not been
carefully examined until recently.  The Sloan Digital Sky Survey
\citep[SDSS;][]{York;2000} with very accurate photometry makes it
possible to investigate the reliability of the SFD map even in those
regions. \citet{Fukugita;2004} tested the region of $\Ebv < 0.15$ in the
SFD map on the basis of number counts of the SDSS DR1
\citep{Abazajian;2003} galaxies, and concluded that the SFD map
prediction is consistent with the number counts.  More recently,
\citet{Schlafly;2010} measured the dust reddening from the displacement
of the bluer edge of the SDSS stellar locus, and found that the SFD map
over-predicts dust reddening by $\sim$ 14\% in $\Ebv$. They also found
that the extinction curve of the Galactic dust is better described by
the \citet{Fitzpatrick;1999} reddening law rather than that of
\citet{O'Donnell;1994}.  These results are also confirmed by an independent
method \citep{Schlafly;2011}.  \citet[][hereafter PG]{Peek;2010}
measured the dust reddening using the passively evolving galaxies as
color standards and found that the SFD map under-predicts reddening where
the dust temperature is low, but at most by 0.045 mag in $\Ebv$.  They
provided the correction map for the SFD with $4^{\circ}.5$ resolution.

A systematic test of the SFD map was also performed by
\citet{Yahata;2007}.  They computed the surface number densities of the
SDSS DR4 \citep{Adelman;2006} photometric galaxies as a function of the
extinction.  They found that the surface number densities of the SDSS
galaxies exhibit a clear positive correlation with the SFD extinction in
the low extinction region, $A_r < 0.1$.  They proposed that the observed
FIR intensity, $I_{100\mum}$, is partially contaminated by the emission of
galaxies along their direction. Since SFD compute the extinction
assuming that the flux is entirely due to the Galactic dust, the region
of more galaxies, therefore with stronger FIR intensity, is assigned a higher
extinction. If the {\it over-estimated} extinction is applied, the
corrected surface number density of galaxies becomes even higher than
the real, resulting in the positive correlation with the extinction as
observed. \citet{Yahata;2007} performed a simple numerical experiment
and showed that even a quite small contamination of FIR emission of
galaxies could qualitatively reproduce the observed anomaly.  Indeed the
expected FIR emission was unambiguously discovered by the subsequent 
stacking image analysis of SDSS galaxies \citep{Kashiwagi;2013}.

The main purpose of the present paper is to reproduce quantitatively the
observed anomaly of the surface number density of SDSS galaxies on the
SFD map by an analytic model of the contamination due to their FIR
emission.

The rest of the paper is organized as follows; after the brief summary
of the SDSS DR7 data \citep{Abazajian;2009} that we use here (\S
\ref{sec:DR7}), we repeat the surface number density analysis of
galaxies introduced by \citet{Yahata;2007}.  Section
\ref{sec:simulation} performs mock numerical simulations so as to
predict the surface number densities of galaxies by taking account of
the effect of their FIR contamination.  We also develop an analytic
model, and make sure that it reproduces well the result of the Mock
simulation in \S \ref{sec:analytic}. The detailed description of our
analytic model is presented in Appendix \ref{app:analytic-detail}.  We
perform the fit to the observed anomaly in the SFD map and find
the mean of the $100\mum$ to $r$-band luminosity ratio, $y=(\nu
L)_{100\mum}/(\nu L)_r$ per SDSS galaxy, is required to be $y_{\rm
avg} > 4$.  Section \ref{sec:discussion} discusses the effect of
the spatial clustering of galaxies, which is neglected either in mock
simulations or in the analytic model.  We also compare the
optimal value of the $100\mum$ to $r$-band flux ratio with that
independently derived with the stacking image analysis by
\citet{Kashiwagi;2013}.  Similar analysis for the corrected SFD map
according to \citet{Peek;2010} is also briefly mentioned.  Finally \S
\ref{sec:conclusions} is devoted to summary and conclusions of the
present paper.

\section{The Sloan Digital Sky Survey DR7}

The SDSS DR7 photometric observation covers 11663 $\rm{deg}^2$ of sky
area, and collects 357 million objects with photometry in five
passbands; $u$, $g$, $r$, $i$, and $z$ \citep[For more details of the
photometric data, see][]{Gunn;1998,Gunn;2006,Fukugita;1996,
Hogg;2001,Ivezic;2004,Smith;2002,Tucker;2006,Padmanabhan;2008,Pier;2003}.
The SDSS photometric data are corrected for the Galactic extinction
according to the SFD map \citep{Stoughton;2002}.  They adopt the
conversion factors from color excess to the dust extinction in each
passband:
%%%%%%%%%%%%%%%%%%%%%%%%%%%%%%%%%%%%%%%
\begin{equation}
k_x \equiv \frac{A_{x,{\SFD}}}{{\Ebv}},
\label{eq:k_x}
\end{equation}
%%%%%%%%%%%%%%%%%%%%%%%%%%%%%%%%%%%%%%%
where $x=u$, $g$, $r$, $i$, and $z$ (Table 6 of SFD).  These factors are
computed assuming the spectral energy density of an elliptical galaxy,
and the reddening law of \citet{O'Donnell;1994} combined with the
extinction curve parameter:
%%%%%%%%%%%%%%%%%%%%%%%%%%%%%%%%%%%%%%%
\begin{equation}
R_V\equiv \frac{A_V}{{\Ebv}}=3.1 .
\label{eq:R_V}
\end{equation}
%%%%%%%%%%%%%%%%%%%%%%%%%%%%%%%%%%%%%%%

The spatial distribution of stellar objects in the SDSS catalogue is
likely to be correlated with the dust distribution.  Therefore the
reliable star-galaxy separation is critical for our present purpose of
testing the SFD map from the distribution of extragalactic objects. We
carefully construct a reliable photometric galaxy sample as follows.

\subsection{Sky area selection}

We choose the regions of SDSS DR7 survey area labeled ``PRIMARY''.
Indeed we found that the ``PRIMARY'' regions in the southern Galactic
hemisphere are slightly different from the area where the objects are
actually located. We are not able to understand why, and thus decide to
use the regions in the northern Galactic hemisphere alone to avoid 
possible problems.

To ensure the quality of good photometric data, we exclude masked
regions.  The SDSS pipeline defines the five types of masked regions
according to the observational conditions.  We remove the four types of
the masked regions, labeled ``BLEEDING'', ``BRIGHT$\_$STAR'', ``TRAIL''
and ``HOLE'' from our analysis.  The masked regions labeled ``SEEING''
is not removed, since relatively bad seeing does not seriously affect
the photometry of relatively bright galaxies that we use in the present
analysis.  The total area of the removed masked regions is about
$340~\rm{deg}^2$, which comprises roughly $4.5\%$ of the entire
``PRIMARY'' regions in the northern Galactic hemisphere.

\subsection{Removing false objects}
\label{subsec:flag-selection}

We remove false objects according to photometry processing flags.  We
first remove fast-moving objects, which are likely the Solar System
objects.  We also discard objects that have bad photometry or were
observed in the poor condition.  A fraction of objects suffers from
deblending problems, {\em i.e.}, the decomposition of photometry images
consisting of superimposed multi-objects is unreliable or failed.  We
remove such objects as well.

\subsection{Magnitude range of galaxies}

The SDSS catalogue defines the type of objects according to the
differences between the {\it cmodel} and PSF magnitudes, where the
former magnitude is computed from the composite flux of the linear
combination of the best-fit exponential and de Vaucouleurs profiles.

Since the reliability of star-galaxy separation depends on the model
magnitude {\it before} extinction correction, we must carefully choose
the magnitude ranges of our sample for the analysis.  In $r$-band, the
star-galaxy separation is known to be reliable for galaxies brighter
than $\sim$21 mag \citep{Yasuda;2001, Stoughton;2002}, while the
saturation of stellar images typically occurs for objects brighter than
15 mag in $r$-band.  Therefore, we choose the magnitude range
conservatively as $17.5 < m_r < 19.4$, where $m_r$ denotes the observed
(extinction uncorrected) magnitudes in $r$-band.  

\begin{figure*}
  \begin{center}
    \includegraphics[width=0.66\textwidth]{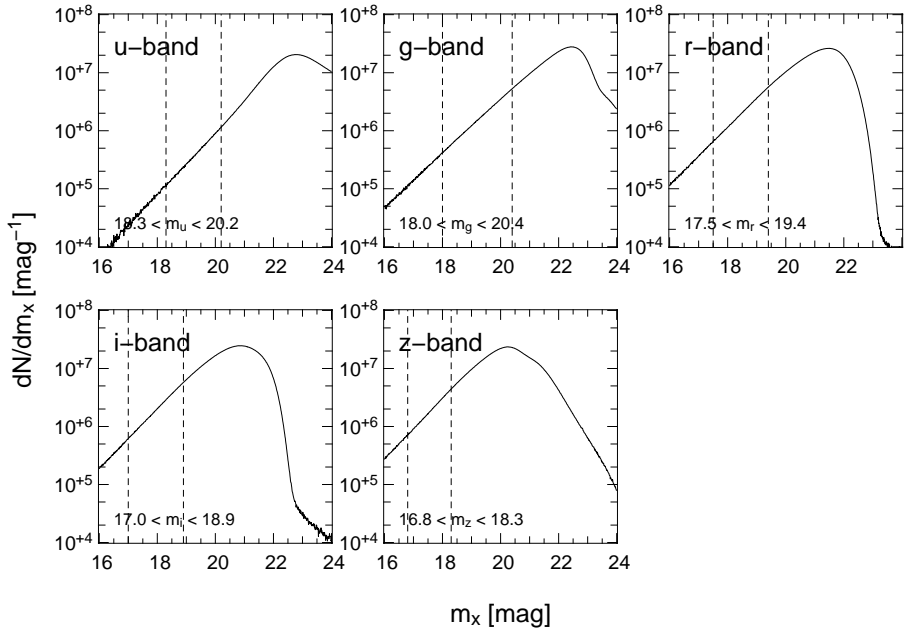}  
 \end{center}
\figcaption{ Differential number counts of the photometric galaxy sample
as functions of extinction uncorrected magnitudes for each band (solid
lines).  The vertical dashed lines indicate the magnitude ranges within
which we use for the analysis. \label{fig:magnitude-distribution}}
\end{figure*}

We adopt the same value of upper/lower limits for extinction corrected
magnitudes.  Figure \ref{fig:magnitude-distribution} shows the
differential number counts of SDSS galaxies as a function of $m_x$ for
each bandpass.  
The faint-end threshold of our $r$-band selected sample, 
$m_r=19.4$,  is $\sim 2$ mag brighter than the turnover of the differential 
number count. We similarly determine the faint-end of magnitude range for 
all bandpasses as 2 mag brighter than the turnover magnitude. 
We confirmed that shifting the upper
or lower limits by $\pm1.0$mag does not significantly change our
conclusions below. We summarize the magnitude range and the number 
of galaxies with and without photometry flag selection for each bandpass in Table \ref{tab:galaxy-number}.

%%%%%%%%%%%%%%%%%%%%%%%%%%%%%%%%%%%%%%%%%%%%%%%%%
%%%%%%%%%%%%%%%%%%%%%%%%%%%%%%%%%%%%%%%%%%%%%%%%%%

%%%%%%%%%%%%%%%%%%%%%%%%%%%%%%%%%%%%%%%%%%%%%%%%%%
\begin{table*}
\caption{The magnitude range and the number of SDSS galaxies for each
bandpass.  The third column shows the number of all SDSS galaxies within
the magnitude range.  The fourth column shows the number of the galaxies
after photometry flag selection described in \S
\ref{subsec:flag-selection}, which are used in our measurement in
\S \ref{sec:DR7}.  The numbers of galaxies are counted without
extinction correction.}  \label{tab:galaxy-number}
\begin{center}
\begin{tabular}{ccccc}
\hline
\hline
bandpass & magnitude range & \# of galaxies & \# of galaxies & rejection rate \\
          &      & (w/o flag selection) &  (w/ flag selection)  &      \\
\hline
$u$ & $18.3 < m_u < 20.2$ & 1200586 & 633319 & 0.472 \\
$g$ & $18.0 < m_g < 20.4$ & 4891030 & 3428064 & 0.299 \\
$r$ & $17.5 < m_r < 19.4$ & 4347881 & 3205638 & 0.263 \\
$i$ & $17.0 < m_i < 18.9$ & 4450724 & 3140684 & 0.295 \\
$z$ & $16.8 < m_z < 18.3$ & 2984104 & 2136639 & 0.284 \\
\hline
\end{tabular}
\end{center}
\end{table*}
%%%%%%%%%%%%%%%%%%%%%%%%%%%%%%%%%%%%%%%%%%%%%%%%%%

%%%%%%%%%%%%%%%%%%%%%%%%%%%%%%%%%%%%%%%%%%%%%%
\section{Surface number densities of SDSS DR7 photometric galaxies} 
\label{sec:DR7}

\subsection{Methodology}

\begin{figure*}
    \begin{center}
    \includegraphics[height=0.33\textwidth]{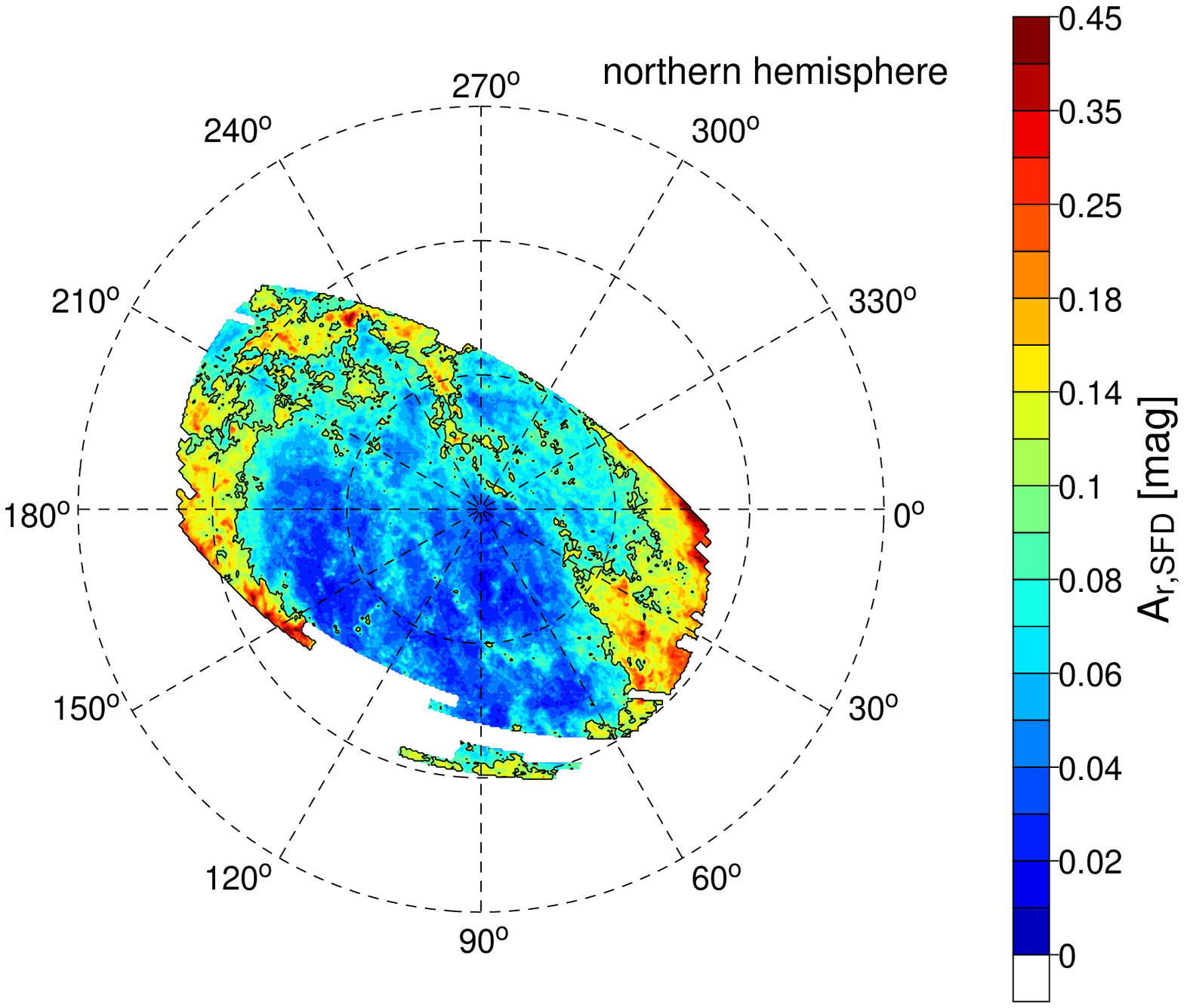}
    \hspace{20pt}
    \includegraphics[height=0.33\textwidth]{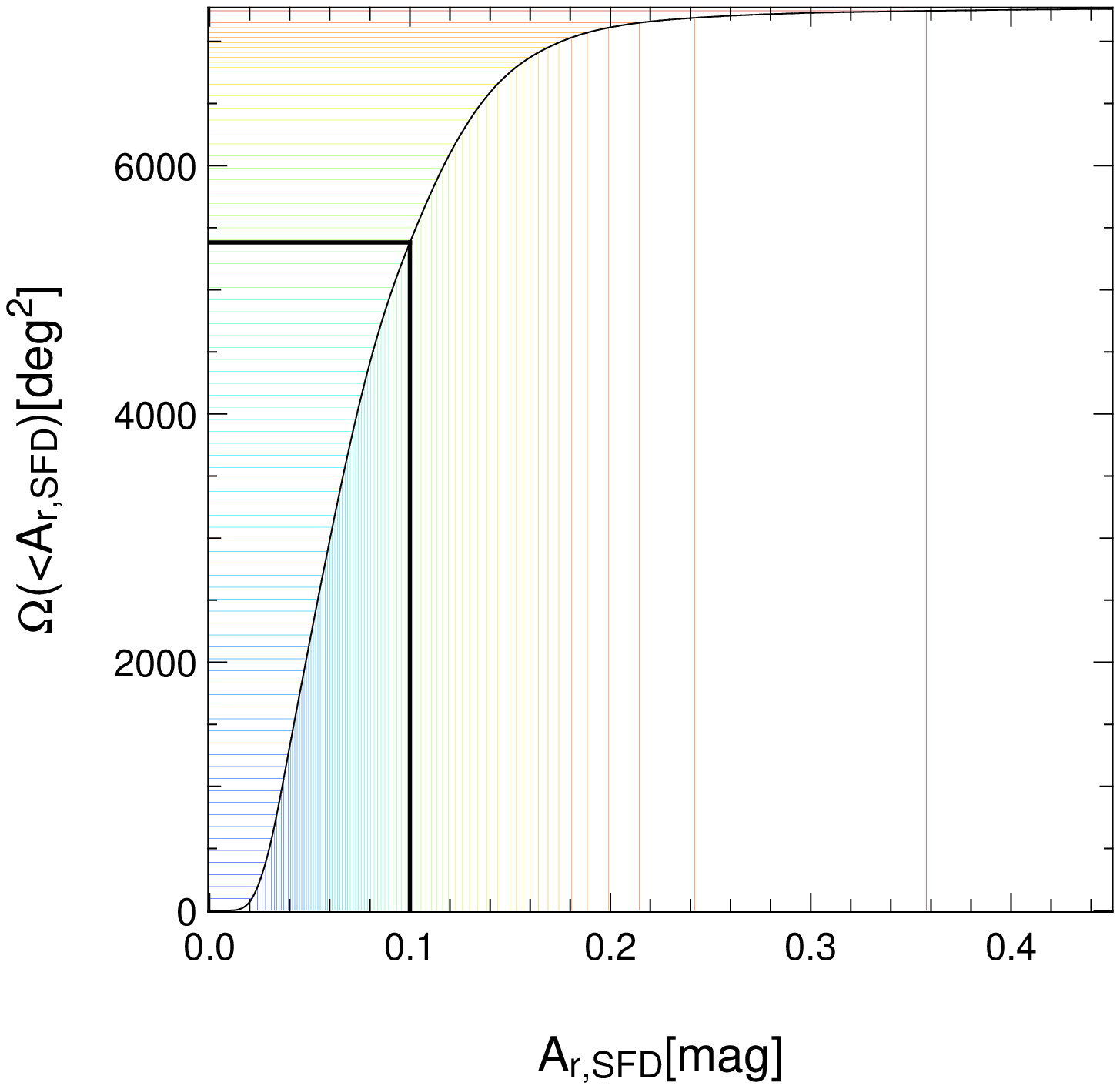}
  \end{center}
\figcaption{Photometric survey area of the SDSS DR7 in Galactic
coordinates ({\it Left}), and the cumulative distribution of the area as
a function of $A_{r,\rm{SFD}}$ ({\it Right}).  The left panel is
color-coded according to the value of $A_{r,\rm{SFD}}$.  The thick lines
in the both panels indicate $A_{r,\rm{SFD}}=0.1$mag, corresponding to
74 $\%$ of the entire survey. The thin lines correspond to each bin of 84
subregions color-coded as the same as the left panel.  \label{fig:survey-region}}
\end{figure*}

In this section, we extend the previous analysis of \citet{Yahata;2007},
and re-examine the anomaly in the surface number density of galaxies
using the SDSS DR7 photometric galaxies, instead of DR4.  The left
panel of Figure \ref{fig:survey-region} plots the sky area of the SDSS
DR7 that is employed in our analysis, where the color scale indicates
the value of the $r$-band extinction provided by SFD, $A_{r,{\SFD}}$.

Since most of the increased survey area of DR7 relative to DR4
corresponds to regions with $A_{r,\mathrm{SFD}}<0.1$mag,
we can study the anomaly in such low-extinction regions discovered by
\citet{Yahata;2007} with higher statistical significance.
%%%%%%%%%%%%%%%%%%%%%%%%%%%%%%%%%%%%%%%%%%%%%%
%%%%%%%%%%%%%%%%%%%%%%%%%%%%%%%%%%%%%%%%%%%%%% 

We first divide the entire sky area of the SDSS DR7 (right panel of
Fig.\ref{fig:survey-region}) into 84 subregions according to the value of 
$A_{r,{\SFD}}$.  Each subregion is chosen so as to have an approximately
same area ($\sim100\rm{deg}^2$), and consists of spatially separated
(disjoint) small patches over the sky.  The right panel of Figure
\ref{fig:survey-region} shows the cumulative area fraction of the sky as
a function of $A_{r,{\SFD}}$.  Note that approximately 74 \% of the
entire sky corresponds to $A_{r,{\SFD}}<0.1$mag, in which
we are interested.
 
Next we count the number of galaxies with the specified range of
$r$-band magnitude in each subregion (\S 2.3), and obtain their surface number
densities as a function of the extinction.  Since the spatial
distribution of galaxies is expected to be homogeneous when averaged
over a sufficiently large area, the surface number densities of galaxies
should be constant, and should not correlate with the extinction. In
other words, any systematic trend with respect to $A_{r,{\SFD}}$ should
indicate to a problem of the SFD map.

\subsection{Results} \label{subsec:results}

\begin{figure*}
  \begin{center}
    \includegraphics[width=0.8\textwidth]{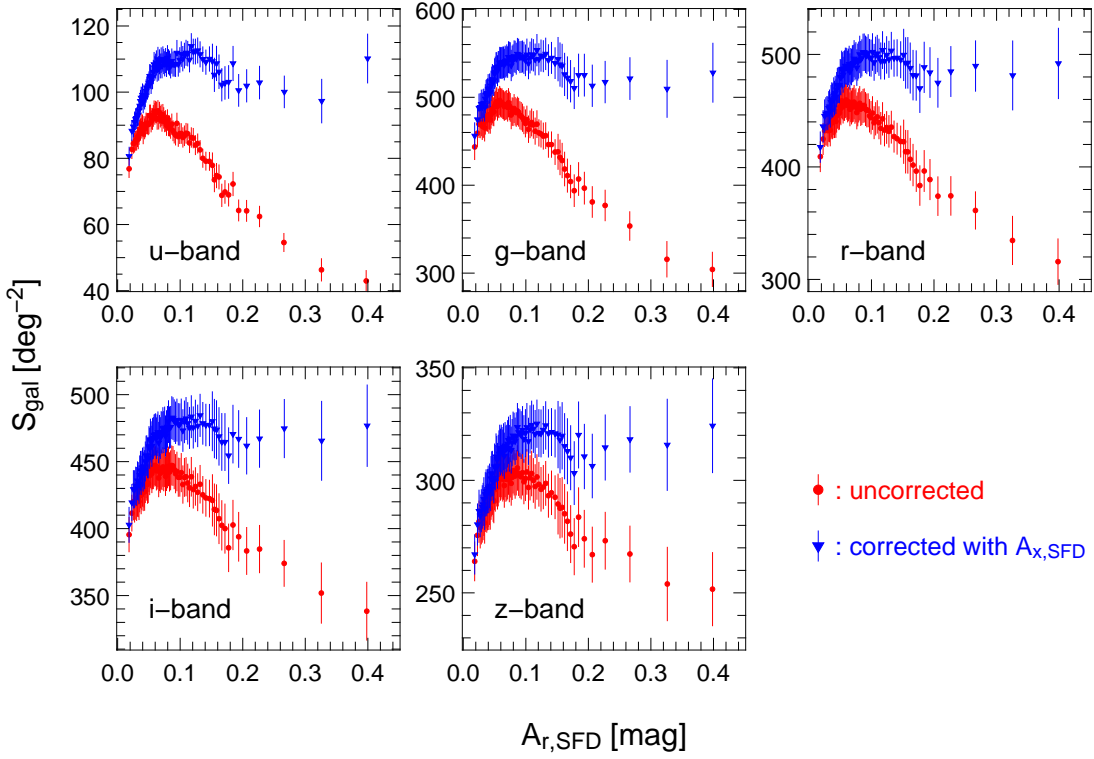}  
 \end{center}
\figcaption{Surface number densities of the SDSS DR7 photometric galaxy
 sample corresponding to Figure \ref{fig:magnitude-distribution},
 against $A_{r,\mathrm{SFD}}$.  The circles/triangles indicates the
 surface number densities calculated with extinction
 un-corrected/corrected magnitudes, respectively.  The statistical
 errors are calculated from equation (\ref{eq:error}).  The horizontal
 axis is the mean of $A_{r,{\SFD}}$ over the galaxies in each subregion.
 \label{fig:photometric-galaxy}}
\end{figure*}

Figure \ref{fig:photometric-galaxy} shows the surface number densities
of galaxies, $S_{\rm{gal}}$, in the 84 subregions for the five
passbands.  The red filled circles indicate $S_{\rm{gal}}$ uncorrected
for dust extinction, while the blue filled triangles are the results
after extinction correction using the SFD map.  Note that the surface
number densities of galaxies in different passpands are plotted against
their corresponding $r$-band extinction, $A_{r,{\SFD}}$.

Following \citet{Yahata;2007} again, we estimate the statistical error
of the surface number density, $\sigma_S^2$, as follows:
%%%%%%%%%%%%%%%%%%%%%%%%%%%%%%%%%%%%
\begin{equation}
\frac{\sigma_S^2}{S^2} = \frac{1}{N} + \frac{1}{\Omega^2} \int_{\Omega}
 \int_{\Omega} w(\theta_{12}) d\Omega_1 d\Omega_2,
\label{eq:error}
\end{equation}
%%%%%%%%%%%%%%%%%%%%%%%%%%%%%%%%%%%%
where $N$ and $S$ denote the number and the surface number density of
the galaxies in the subregion of area $\Omega$, and $w(\theta_{12})$ is
the angular correlation function of galaxies with $\theta_{12}$ being
the angular separation between two solid angle elements, $d\Omega_1$ and
$d\Omega_2$.  The first term in equation (\ref{eq:error}) denotes the
Poisson noise, while the second term comes from galaxy clustering. 

 For definiteness, we adopt the double power-law model
\citep{Scranton;2002, Fukugita;2004} for $w(\theta_{12})$:
%%%%%%%%%%%%%%%%%%%%%%%%%%%%%%%%%%%%
\begin{equation}
w(\theta_{12}) = \cases{
0.008(\theta_{12}/\rm{deg})^{-0.75} & $(\theta_{12} \leq 1 \rm{deg})$ \cr
0.008(\theta_{12}/\rm{deg})^{-2.1} & $(\theta_{12} > 1 \rm{deg})$ \cr
}.	
\label{eq:angular-2PCF}
\end{equation}
%%%%%%%%%%%%%%%%%%%%%%%%%%%%%%%%%%%%
Strictly speaking, the integration in the second term of equation
(\ref{eq:error}) should be performed over a complex and disjoint shape
of each subregion.  For simplicity, however, we substitute the
integration over a circular region whose area is equal to that of the
actual subregion.  Although this approximation may overestimate
the true error, it does not affect our conclusion at all.  For the
typical values of $\Omega\sim 100\rm{deg}^2$ and
$S\sim480\rm{deg}^{-2}$, we find that the second term is larger by two
orders of magnitude than the first Poisson-noise term.

Figure \ref{fig:photometric-galaxy} suggests that the SFD correction
works well in relatively high-extinction regions, {\it i.e.,}
$A_{r,{\SFD}}>0.1$; before corrected for extinction, the surface number
density of galaxy, $S_{\rm{gal}}$, monotonically decreases against
$A_{r,{\SFD}}$ as naturally expected. It becomes roughly constant within
the statistical error after extinction correction.

In low-extinction regions ($A_{r,{\SFD}} < 0.1$), however, the
uncorrected $S_{\rm{gal}}$ {\it increases} with $A_{r,{\SFD}}$, which is
opposite to the behavior expected from the Galactic dust extinction.
The anomalous positive correlation between surface number densities and
extinction is even more enhanced after the extinction correction.  Apart
from the slight quantitative differences, these results are consistent
with the trend discovered for the SDSS DR4 by \citet{Yahata;2007},
especially for the positive correlations in $A_{r,\rm SFD}<0.1$.

\citet{Yahata;2007} argued that the trend is due to the presence
of the FIR emission of galaxies, which contaminates the 100 {\mum} flux
of IRAS that is conventionally ascribed to the Galactic dust entirely.
Indeed their hypothesis is now directly confirmed by the stacking
analysis of \citet{Kashiwagi;2013}, who detected the unambiguous
signature of FIR emission from SDSS galaxies in the SFD map. Our next task, therefore,
is to ask if the detected nature of the FIR emission of galaxies by
\citet{Kashiwagi;2013} properly accounts for the anomaly that we
described here. In what follows, we consider the surface number density
of the galaxies measured in $r$-band alone, simply because it is the
central SDSS passband, and the result is equally applicable to the other
passbands.

%%%%%%%%%%%%%%%%%%%%%%%%%%%%%%%%%%%%%%%%%%%%%%%%%
%%%%%%%%%%%%%%%%%%%%%%%%%%%%%%%%%%%%%%%%%%%%%%%%%

%%%%%%%%%%%%%%%%%%%%%%%%%%%%%%%%%%%%%%%%%%%%%%%%%
\section{Mock numerical simulation to compute 
 the FIR contamination effect of galaxies on the extinction map}
\label{sec:simulation}
%%%%%%%%%%%%%%%%%%%%%%%%%%%%%%%%%%%%%%%%%%%%%%%%%

In this section, we present the results of mock numerical simulations
that take into account the effect of the FIR emission of mock galaxies
in a fairly straightforward manner.  First we randomly place mock
galaxies over the SDSS DR7 sky area so that they have the same number
density and the same $r$-band magnitude distribution of the SDSS DR7
sample. Next, we assign a $100{\mum}$ flux to each mock galaxy according
to the probability distribution function discussed in \S
\ref{subsec:IRAS_SDSS}. We sum up the $100{\mum}$ fluxes of the mock
galaxies over the raw SFD map that is assumed to be {\it not
contaminated} by the FIR emission of mock galaxies, and construct a {\it
contaminated} mock extinction map. Finally, we compute the surface
number densities of mock galaxies exactly as we did for the real galaxy
sample.  Further details are described below.

%%%%%%%%%%%%%%%%%%%%%%%%%%%%%%%%%%%%%%%%%%%%%%%%%
\subsection{Empirical correlation between 100$\mu$m and r-band luminosities 
of PSCz/SDSS galaxies} \label{subsec:IRAS_SDSS}

 In order to assign 100{\mum} emission to each mock galaxy with a given
$r$-band magnitude, we need an empirical relation between the two
luminosities, $L_{100\mum}$ and $L_{\rm r}$. For that purpose,
\citet{Yahata;thesis} created a sample of galaxies detected both in SDSS
and in PSCz \citep[IRAS Point Source Catalog Redshift
Survey;][]{Saunders;2000}. To be more specific, he searches for SDSS
galaxies within 2 arcmin from the position of each PSCz galaxy, and
selects the brightest one as the optical counterpart.  Approximately
95\% of the PSCz galaxies within the SDSS survey region have SDSS
counterparts, and the resulting sample consists of 3304 galaxies in total.
Note, however, that the sample is biased towards the FIR luminous
galaxies since SDSS optical magnitude-limit is significantly deeper than
that of PSCz galaxies.

%%%%%%%%%%%%%%%%%%%%%%%%%%%%%%%%%%%%%%%%%%%%%%%%%%%
\begin{figure*}
  \begin{center}
    \includegraphics[height=0.38\textwidth]{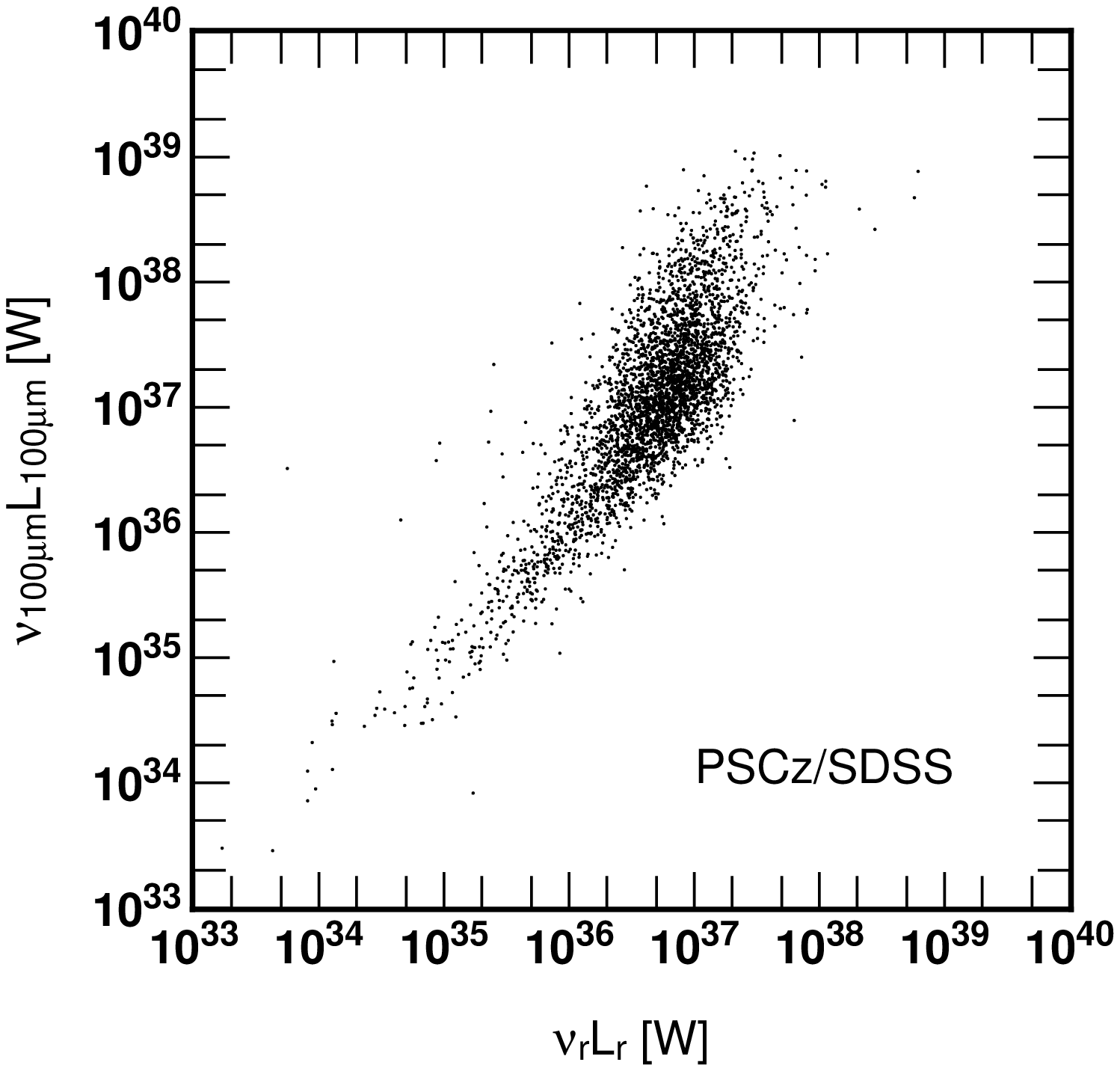}
    \hspace{5mm}
     \includegraphics[height=0.38\textwidth]{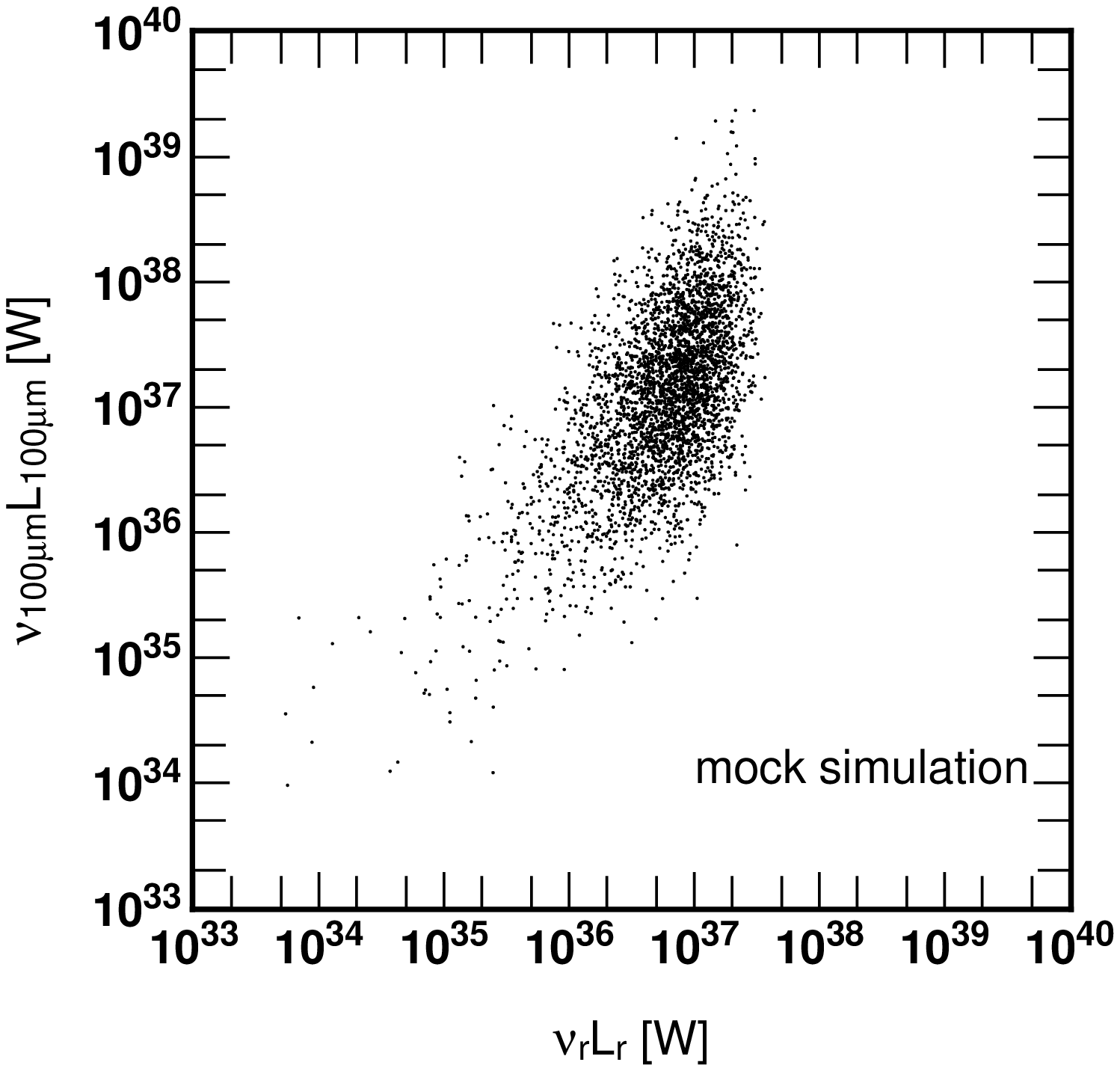}
    \end{center}
\figcaption{ {\em left panel}; Relation between $\nu_{100\mu \rm{m}}
L_{100\mu\rm{m}}$ and $\nu_r L_r$ for the PSCz/SDSS overlapped galaxies.
{\em right panel}; same as the left panel, but for the mock galaxies 
generated based on $r$-band luminosity function (equation \ref{eq:LFopt}), 
the log-normal PDF of $y$ adopting the parameters in equation (\ref{eq:y-value-entire}), 
and the flux cut $f_{\rm 100\mum}<1.0{\rm Jy}$.
\label{fig:nuL-distribution}}
\end{figure*}
%%%%%%%%%%%%%%%%%%%%%%%%%%%%%%%%%%%%%%%%%%%%%%%%%%%

The left panel of Figure \ref{fig:nuL-distribution} shows the relation
between $\nu_{100\mu\rm{m}}L_{100\mu\rm{m}}$ (PSCz) and $\nu_r L_r$
(SDSS) of the PSCz/SDSS overlapped sample.  For K-correction, we use the
``K-corrections calculator'' service \citep{Chilingarian;2010} for
$r$-band, and extrapolate the FIR flux at 100$\mum$ from the 
second-order polynomials using 25 and 60 {\mum} fluxes 
\citep{Takeuchi;2003}.  

%%%%%%%%%%%%%%%%%%%%%%%%%%%%%%%%%%%%%%%%%%%%%%%%%%%
\begin{figure*}
  \begin{center}
    \includegraphics[height=0.4\textwidth]{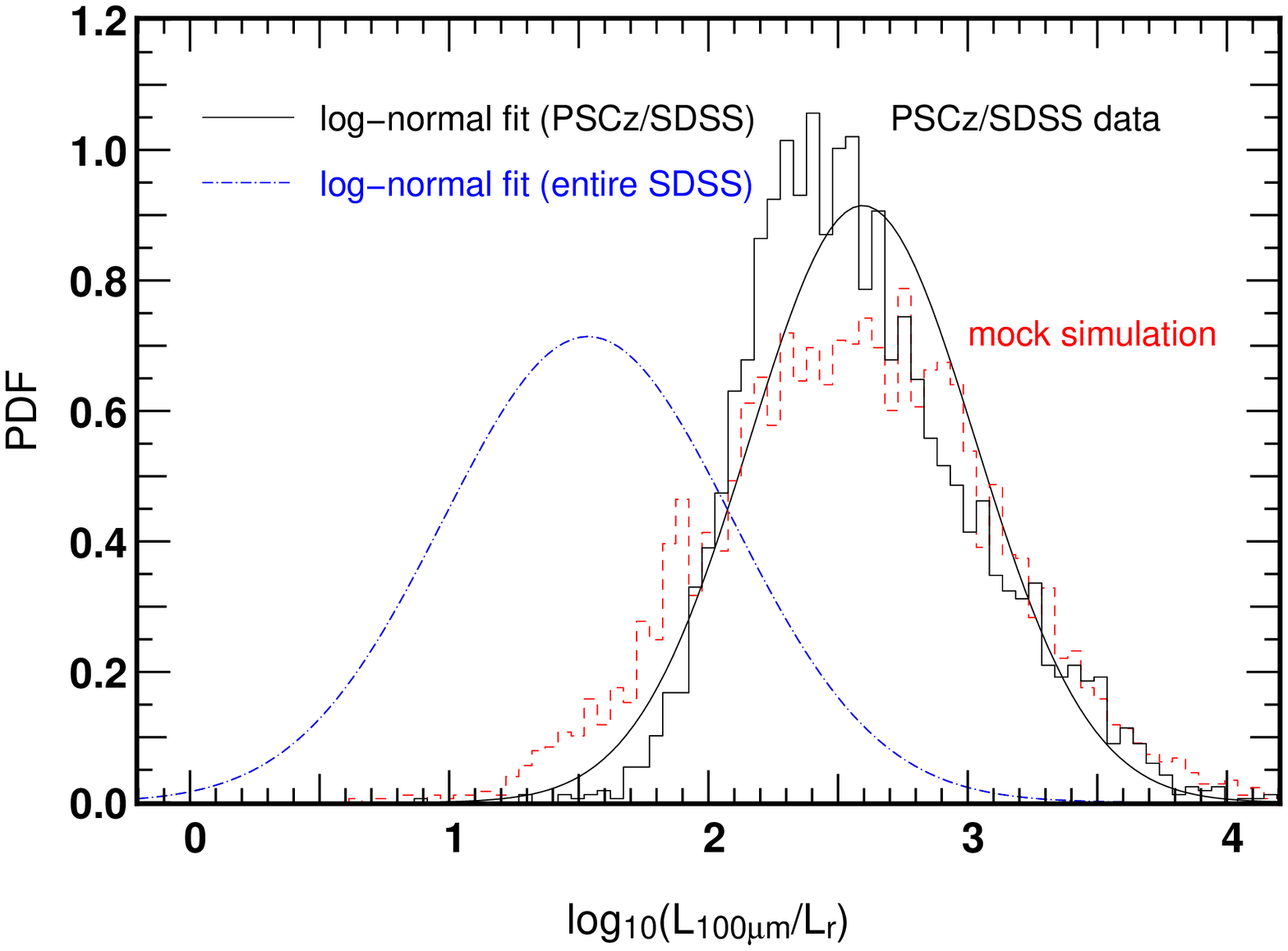}
    \end{center}
\figcaption{The probability distribution function of $L_{100\mu
\rm{m}}/L_r$; the PSCz/SDSS overlapped sample (histogram), 
the best-fit log-normal function (black solid curve), 
flux-limited mock galaxies (red dashed histogram),
and the best-fit log-normal function estimated for the entire SDSS 
galaxies (blue dot dashed curve).  \label{fig:histoLL}}
\end{figure*}
%%%%%%%%%%%%%%%%%%%%%%%%%%%%%%%%%%%%%%%%%%%%%%%%%%%

The resulting scatter plot indicates that $L_{100\mum}$ and $L_r$ are
approximately proportional, albeit with considerable scatter.  So we
compute the probability distribution function (PDF) of the luminosity
ratio, 
%%%%%%%%%%%%%%%%%%%%%%%%%%%%%%%%%%%
\begin{equation}
y \equiv \frac{\nu_{100\mu\rm{m}}L_{100\mu \rm{m}}}{\nu_r L_r},
\end{equation}
%%%%%%%%%%%%%%%%%%%%%%%%%%%%%%%%%%%
for the sample (solid histogram in Figure \ref{fig:histoLL}), and find that the PDF is
reasonably well described by a log-normal distribution:
%%%%%%%%%%%%%%%%%%%%%%%%%%%%%%%%%%%
\begin{equation}
P_{\mathrm{ratio}}(y)dy = \frac{1}{y \ln 10 \sqrt{2\pi \sigma^2}}
\exp \left[ -\frac{(\log_{10}y-\mu)^2}{2\sigma^2}\right] dy,
\label{eq:log-normal}
\end{equation}
%%%%%%%%%%%%%%%%%%%%%%%%%%%%%%%%%%%
%where $\mu = 2.591$ and $\sigma =0.428$ are the mean and dispersion of
where $\mu = 0.393$ and $\sigma =0.428$ are the mean and dispersion of
$\log_{10} y$ (solid curve in Figure \ref{fig:histoLL}).

Since the PSCz/SDSS overlapped sample is a biased sample in a sense that
these galaxies are selected towards the FIR luminous galaxies, the above
log-normal distribution is not necessarily applicable for the entire SDSS
galaxies.  Therefore we assume the FIR-optical luminosity ratio of the
{\it entire} SDSS galaxies also follows a log-normal distribution, and
estimate the values of $\mu$ and $\sigma$ for the entire sample by considering 
the PSCz detection limit.  Although the flux limit of PSCz is defined through
$f_{60{\mum}}>0.6\rm{Jy}$, we roughly estimate the corresponding
effective flux limit at 100{\mum} is $f_{100{\mum}}>1.0\rm{Jy}$ from the
distribution of $f_{100{\mum}}$ for the PSCz/SDSS galaxies (Left-panel
of Figure \ref{fig:nuL-distribution}).

Armed with these assumptions, the number of the galaxies that are
detected by this flux cut and have the luminosity between
$L_r\sim L_r+dL_r$ and $L_{100{\mum}}\sim L_{100{\mum}}+dL_{100{\mum}}$
is calculated as,
%%%%%%%%%%%%%%%%%%%%%%%%%%%%%%%%%%%
\begin{eqnarray}
N^{\rm{obs}}&(&L_r, L_{100{\mum}})dL_r dL_{100{\mum}} \cr
&=&
 \frac{\Omega_s}{4\pi} \bigg[ \int^{\infty}_0 dz \frac{dV(<z)}{dz}
 \Theta(L_{100{\mum}},z) \cr
&&\times
\Phi(L_r)
  P(L_{100{\mum}}|L_r;\mu,\sigma) \bigg] dL_r dL_{100{\mum}},
\label{eq:Nobs}
\end{eqnarray}
%%%%%%%%%%%%%%%%%%%%%%%%%%%%%%%%%%%
where $\Omega_s$ is the solid angle of the PSCz/SDSS overlapped survey
area, and $V(<z)$ denotes the co-moving volume up to redshift $z$. 
The step function $\Theta(L_{100{\mum}},z)$ describes the flux cut of
PSCz:
%%%%%%%%%%%%%%%%%%%%%%%%%%%%%%%%%%%%
\begin{equation}
\Theta(L_{100{\mum}},z) = \cases{
	1 & $(L_{100{\mum}}/4\pi d^2_L(z)>1.0{\rm Jy}$) \cr
	0 & $(\rm{else})$ \cr
},	
\label{eq:ThetaL100}
\end{equation}
%%%%%%%%%%%%%%%%%%%%%%%%%%%%%%%%%%%%
where $d_L(z)$ is the luminosity distance at redshift $z$.

 We adopt the double-Schechter luminosity function in $r$-band measured
from the SDSS DR2 data \citep{Blanton;2005} for $\Phi(L_r)$:
%%%%%%%%%%%%%%%%%%%%%%%%%%%%%%%%%%%%%%%%%
\begin{eqnarray}
\Phi (L_r) dL_r
&=&
\frac{dL_r}{L_{r,\ast}} \exp \left( -\frac{L_r}{L_{r,\ast}}\right) \cr
&\times&
\left[ \phi_{\ast,1} \left(\frac{L_r}{L_{r,\ast}}\right)^{\alpha_1}
 + \phi_{\ast,2}\left(\frac{L_r}{L_{r,\ast}}\right)^{\alpha_2}\right].
\label{eq:LFopt}
\end{eqnarray}
%%%%%%%%%%%%%%%%%%%%%%%%%%%%%%%%%%%%%%%%%%%%
The conditional probability density function of $L_{100{\mum}}$ for
given $L_r$ is assumed to be log-normal:
%%%%%%%%%%%%%%%%%%%%%%%%%%%%%%%%%%%%
\begin{eqnarray}
&&P(L_{100{\mum}}|L_r;\mu,\sigma)dL_{100{\mum}}
=\frac{1}{\ln 10\sqrt{2\pi \sigma^2}} \cr
&\times&
\exp\left(-\frac{[\log (\nu_{100\mum}L_{100{\mum}}/\nu_r L_r)-\mu]^2}{2\sigma^2}
\right)
\frac{dL_{100{\mum}}}{L_{100{\mum}}} \cr
&=& yP_{\rm ratio}(y;\mu,\sigma)\frac{dL_{100\mum}}{L_{100\mum}}.
\label{eq:conditional}
\end{eqnarray}
%%%%%%%%%%%%%%%%%%%%%%%%%%%%%%%%%%%%

We use equation (\ref{eq:Nobs}) to find the best-fit $\mu$ and $\sigma$
in equation (\ref{eq:conditional}) for the {\it entire} SDSS galaxies
that reproduce the observed distribution of the PSCz/SDSS overlapped sample. 
The resulting values are $\mu=-0.662$ and $\sigma=0.559$ as 
plotted in blue dot-dashed line in Figure
\ref{fig:histoLL}.  This result indicates that the mean value of $y$ of 
the PSCz/SDSS overlapped sample is biased by
an order of magnitude relative to that for the entire galaxies; 
see equation (\ref{eq:y-value-pscz}) and (\ref{eq:y-value-entire}).  

\begin{figure*}
  \begin{center}
     \includegraphics[height=0.32\textwidth]{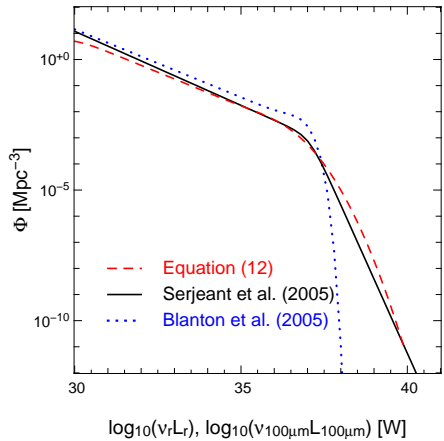}
    \end{center}
\figcaption{Luminosity function (LF) of galaxies 
at $100\mum$ and $r$-band. Solid line is 100$\mum$ LF directly measured 
from the PSCz data \citep{Serjeant;2005}, while dashed line shows our estimate of 100$\mum$ 
LF based on equation (\ref{eq:LF_FIR}) with the best-fit $\mu$, $\sigma$ and $r$-band LF 
\citep[blue dotted line]{Blanton;2005}. \label{fig:LF}}
\end{figure*}

Adopting now the best-fit log-normal distribution, the luminosity
function at $100{\mum}$ is calculated as
%%%%%%%%%%%%%%%%%%%%%%%%%%%%%%%%%%%
\begin{equation}
\label{eq:LF_FIR}
\Phi(L_{100{\mum}}) = \int^{\infty}_0 dL_r\Phi(L_r) 
P(L_{100{\mum}}|L_r;\mu,\sigma).
\end{equation}
%%%%%%%%%%%%%%%%%%%%%%%%%%%%%%%%%%%
As plotted in Figure \ref{fig:LF}, the above best-fit indeed agrees well
with the luminosity function independently measured from the PSCz data
\citep{Serjeant;2005}.

%%%%%%%%%%%%%%%%%%%%%%%%%%%%%%%%%%%%%%%%%%%%%%%%%%%
%%%%%%%%%%%%%%%%%%%%%%%%%%%%%%%%%%%%%%%%%%%%%%%%%%%

 In order to make sure if the above FIR log-normal PDF combined with the
FIR flux cut reproduces the left panel of Figure
\ref{fig:nuL-distribution}, we generate mock galaxies and assign $z$,
$L_r$, and $L_{100{\mum}}$ following the redshift distribution $dV(<z)$,
and equations (\ref{eq:LFopt}) and (\ref{eq:conditional}).  Then we
exclude those mock galaxies with $f_{100{\mum}}<1.0\rm{Jy}$ to mimic the
flux cut.  The right panel of Figure \ref{fig:nuL-distribution} and the
dashed histogram in Figure \ref{fig:histoLL} show the
resulting luminosity distribution and the PDF of $y$ for those
mock galaxies.  Although not perfect, the mock galaxies reproduce the observed
distribution reasonably well. We suspect that the discrepancy between
the observed data and the mock simulation is mainly due to the
limitation of our log-normal approximation neglecting the dependence of
the ratio $L_{100{\mum}}/L_{60{\mum}}$ on $L_{100{\mum}}$.

 For simplicity of the procedure, however, we adopt the best-fit
log-normal distribution as the fiducial model of the 100 {\mum} flux of
the SDSS galaxies in what follows.  In doing so, we parametrize the
distribution by $y_{\rm avg}$ and $y_{\rm rms}$ instead of $\mu$ and
$\sigma$:
%%%%%%%%%%%%%%%%%%%%%%%%%%%%%%%%%%%%%%%%%%%%%%%
\begin{eqnarray}
y_{\rm avg} &=& e^{\mu\ln 10+ (\sigma \ln 10)^2/2}, \\
y_{\rm rms} &=& e^{\mu \ln 10 +(\sigma \ln 10)^2/2}
\sqrt{e^{(\sigma \ln 10)^2}-1} ,
\end{eqnarray}
%%%%%%%%%%%%%%%%%%%%%%%%%%%%%%%%%%%%%%%%%%%%%%%
since the anomaly is basically determined by $y_{\rm avg}$ as will be
shown in Figure \ref{fig:various-parameters} below. For definiteness,
the PSCz/SDSS overlapped sample is characterized by
%%%%%%%%%%%%%%%%%%%%%%%%%%%%
\begin{equation}
\mu=0.393, \sigma=0.428, y_{\rm avg}=4.015, y_{\rm rms}=5.143, 
\label{eq:y-value-pscz}
\end{equation}
%%%%%%%%%%%%%%%%%%%%%%%%%%%%
while the entire SDSS sample is estimated to have
%%%%%%%%%%%%%%%%%%%%%%%%%%%%
\begin{equation}
\mu=-0.662, \sigma=0.559, y_{\rm avg}=0.499, y_{\rm rms}=1.026.
\label{eq:y-value-entire}
\end{equation}
%%%%%%%%%%%%%%%%%%%%%%%%%%%%

%%%%%%%%%%%%%%%%%%%%%%%%%%%%%%%%%%%%%%%%%%%%%%
\subsection{Simulations \label{subsec:poisson}}

Now we are in a position to present our mock simulations that exhibit 
the effect of the FIR contamination of galaxies.  In this
subsection, we neglect the spatial clustering of galaxies and consider
the case for Poisson distributed mock galaxies. The effect of spatial
clustering of galaxies will be discussed separately in \S
\ref{subsec:N-body}.  Our mock simulations are performed as follows.
%%%%%%%%%%%%%%%
\begin{enumerate}
 \item We distribute random particles as mock galaxies over the
       SDSS DR7 survey area. The number of the mock galaxies is
       adjusted so as to approximately match that of the SDSS photometric galaxies.
 \item We assign an intrinsic apparent magnitude in
 $r$-band to each mock galaxy so that the resulting magnitude
 distribution reproduces that of the SDSS galaxies (Figure
       \ref{fig:magnitude-distribution}).
\item Assign $100\mu\mathrm{m}$ flux to each mock galaxy adopting the
       log-normal PDF for the $100\mum$-to-$r$-band flux ratio, $y$. The
      PDF is characterized by $y_{\rm avg}$ and $y_{\rm rms}$.
\item We convolve the $100{\mum}$ fluxes of the mock galaxies
       with a $\rm{FWHM} = 5'.2$ Gaussian filter, so as 
       to mimic the SFD resolution, $\rm{FWHM} = 6'.1$ 
       (see also Appendix \ref{app:psf}). Those mock galaxies with 100 {\mum}
       flux being larger than $1.0\rm{Jy}$ are excluded, since SFD individually
       subtracted the 100{\mum} emission of those bright galaxies. 
       We include only the contribution of the mock galaxies with 
       $17.5<m_{\rm r}<19.4$ so as to be consistent with our analysis in \S
       \ref{subsec:results}. We note, however, that in reality the FIR
       contamination would be likely contributed by galaxies outside the
       magnitude range (not only SDSS galaxies but non-SDSS galaxies
       that do not satisfy the SDSS selection criteria). Therefore the
       current mock simulation should be interpreted to see the extent to
       which the SDSS galaxies in that magnitude range alone account 
       for the observed anomaly in their surface number density.
\item We superimpose the 100{\mum} intensity of the mock galaxies
       on a true extinction map and construct a contaminated extinction
       map after subtracting the background ({\it i.e.,} mean) level of
       the mock galaxy emission.  In what follows, the resulting
       extinction with mock galaxy contaminated is denoted as
       $A_r^{\prime}$.
\item Finally, we calculate $S_{\rm{mock}}$, surface number densities of
       mock galaxies whose corrected/uncorrected magnitudes lie between
       17.5 and 19.4 mag, repeating the same procedure discussed in
       \S\ref{sec:DR7}, but using $A_r^{\prime}$ instead.
\end{enumerate}
%%%%%%%%%%%%%%%%%%%%%%%%%%

 Note that our mock analysis uses the SFD map as the true extinction map
without being contaminated by FIR emission of {\it mock} galaxies. Of course,
the SFD map is contaminated by FIR emission from {\it real}
galaxies, and thus cannot be regarded as a {\em true} extinction map for
them. Nevertheless the contamination of real galaxies should not be
correlated at all with the mock galaxies. This is why the SFD map can be
used as the true extinction map for the current simulation.

The {\it observed} magnitude of each mock galaxy, {\it i.e.,} affected
by the Galactic dust absorption alone, is calculated from the true, in
the present case the SFD map, but the extinction correction is done using
$A^{\prime}_r$.  Note that the difference between the true map and the
contaminated map affects the value of extinction of regions where mock
galaxies are located. Therefore, surface number densities of mock
galaxies {\it before} the extinction correction are also influenced by the
FIR contamination.

\begin{figure*}
  \begin{center}
     \includegraphics[width=0.33\textwidth]{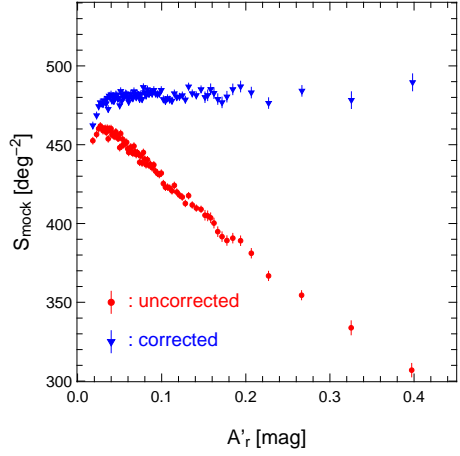}	
  \end{center}
\figcaption{The surface number densities of the randomly distributed
mock galaxies with assigned magnitude of $17.5 < m_r < 19.4$.  
The symbols are the same as in
Figure \ref{fig:photometric-galaxy}.  The values of $y_{\rm avg}$ and
$y_{\rm rms}$ estimated for the entire SDSS galaxies are adopted,
instead of those for the PSCz/SDSS overlapped sample.  The error bars
reflect the Poisson noise alone.  \label{fig:fiducial-parameters}}
\end{figure*}

 Figure \ref{fig:fiducial-parameters} shows the surface number densities
of mock galaxies as a function of $A_r^{\prime}$.  Here we adopt $y_{\rm
avg}=0.499$ and $y_{\rm rms}=1.026$, i.e., equation (\ref{eq:y-value-entire}) 
which are estimated for the entire SDSS galaxy sample.  
The quoted error bars in the panel reflect the
Poisson noise alone.  The results exhibit a similar, but significantly
weak correlation with $A_{r,{\rm SFD}}$ at $A_{r,{\rm SFD}}<0.1$ compared to the observed one 
(Fig.\ref{fig:photometric-galaxy}), especially for the extinction-uncorrected surface densities.

%%%%%%%%%%%%%%%%%%%%%%%%%%%%%%%%%%%%%%%%%%%%%%%%
%%%%%%%%%%%%%%%%%%%%%%%%%%%%%%%%%%%%%%%%%%%%%%%%%

\begin{figure*}
  \begin{center}
    \includegraphics[height=0.4\textheight]{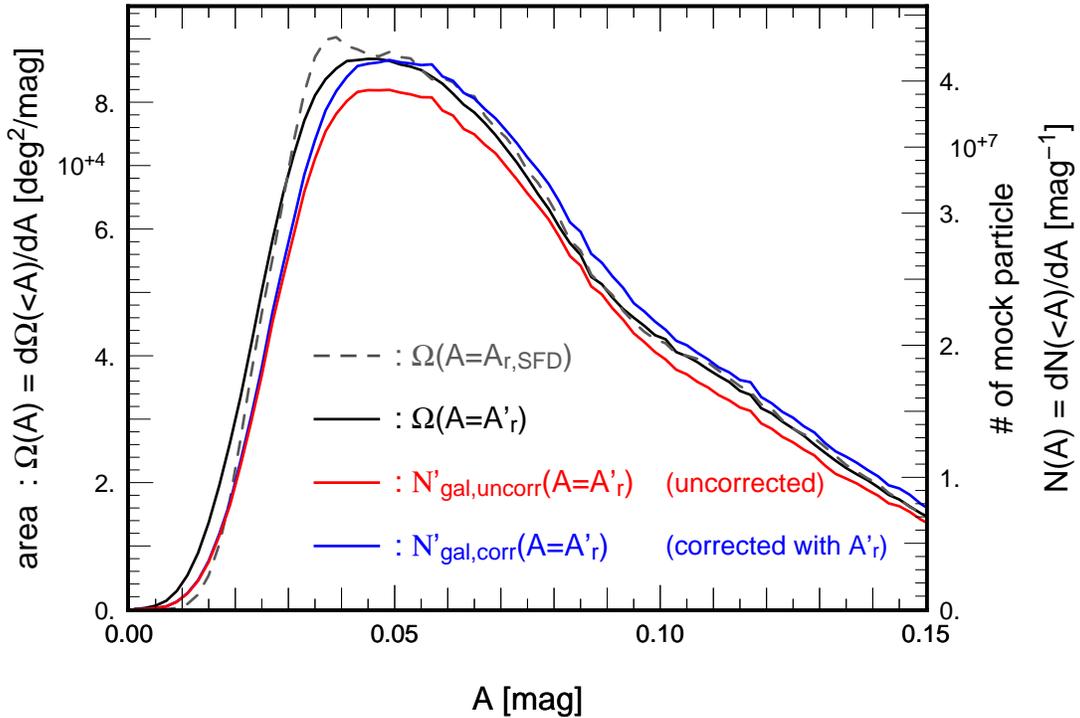}
  \end{center}
\figcaption{The distribution of sky area and mock galaxies.  The dashed
line is the distribution of sky area as a function of {\em true}
extinction, $A$, and the solid black line is calculated as a function of
contaminated extinction, $A+\Delta A$.  The red (blue) line indicates
the distribution of number of galaxies as a function of contaminated
extinction, $A+\Delta A$, with uncorrected (corrected) using the
contaminated extinction.  The distributions of number of galaxies are
divided by the average surface number density, therefore surface number
densities are equal to the average at the points where the distribution
of sky area and number of galaxies cross. We have adopted $y_{\rm avg}=10$ 
and $y_{\rm rms}=5$ for clear visualization of the differences between each lines.
\label{fig:distribution}}
\end{figure*}

Figure \ref{fig:distribution} would help us to understand the origin of
the anomaly intuitively. (In this plot, we have adopted $y_{\rm avg}=10$ 
and $y_{\rm rms}=5$ just to clearly visualize the trends discussed in the following.) 
The dashed line indicates the differential
distribution of the sky area as a function of $A_{r,\mathrm{SFD}}$,
$\Omega(A_{r,\mathrm{SFD}})$, which corresponds to the derivative of the
left panel of Figure \ref{fig:survey-region}.  The black solid line
shows the same distribution, but as a function of $A_r^{\prime}$.  The
resulting $\Omega^{\prime}(A^{\prime}_r)$ slightly differs from
$\Omega(A_{r,\mathrm{SFD}})$ due to the FIR contamination of mock
galaxies.

The blue and red solid lines in Figure \ref{fig:distribution} show the
differential number counts of galaxies, $N^{\prime}_{\rm gal,uncorr}$ and
$N^{\prime}_{\rm gal,corr}$, as a function of $A_r^{\prime}$
calculated from magnitudes uncorrected/corrected for extinction with $A'_r$.
 The shapes of $N^{\prime}_{\rm gal,uncorr}$ and
$N^{\prime}_{\rm gal, corr}$ are slightly shifted towards
the right relative to $\Omega^{\prime}(A_r^{\prime})$, because the
pixels with more galaxies suffer from the larger contamination and thus
have larger values of $A_r^{\prime}$.

Although the amount of this shift is quite small on average, the
differences between $\Omega^{\prime}$ and the differential number counts
for the same $A_r^{\prime}$ become larger in low-extinction regions
because $\Omega^{\prime}$ is a rapidly increasing function of
$A_r^{\prime}$.  Therefore the surface number densities,
$N^{\prime}_{\rm gal,uncorr}$ or $N^{\prime}_{\rm gal,corr}$ divided
by $\Omega^{\prime}$, drastically change especially in low-extinction
regions. In other words, the correlation between the surface number
densities and $A_r^{\prime}$ is significantly enhanced due to the nature
of the SDSS sky area and the SFD map. This also implies that the shape
of the anomaly in $S_{\rm gal}$ is basically determined by the functional form 
of $\Omega(<A)$.
%%%%%%%%%%%%%%%%%%%%%%%%%%%%%%%%%%%%%%%%%%%%%%%%%
%%%%%%%%%%%%%%%%%%%%%%%%%%%%%%%%%%%%%%%%%%%%%%%%%

We also investigate how this result is affected by the $100\mum$
emission of galaxies outside the magnitude range.  We incorporate the
$100{\mum}$ flux of mock galaxies within a wider magnitude range ($15.0
< m_r < 21.0$), but the result is almost indistinguishable.  This is
mainly because that the additional contamination is not directly
correlated with the surface number densities that we measure, partly
because we neglect spatial clustering of galaxies.  Therefore it affects
only as the statistical noise in the extinction map, and does not
contribute to the systematic correlation.

\begin{figure*}
  \begin{center}
    \includegraphics[width=0.6\textwidth]{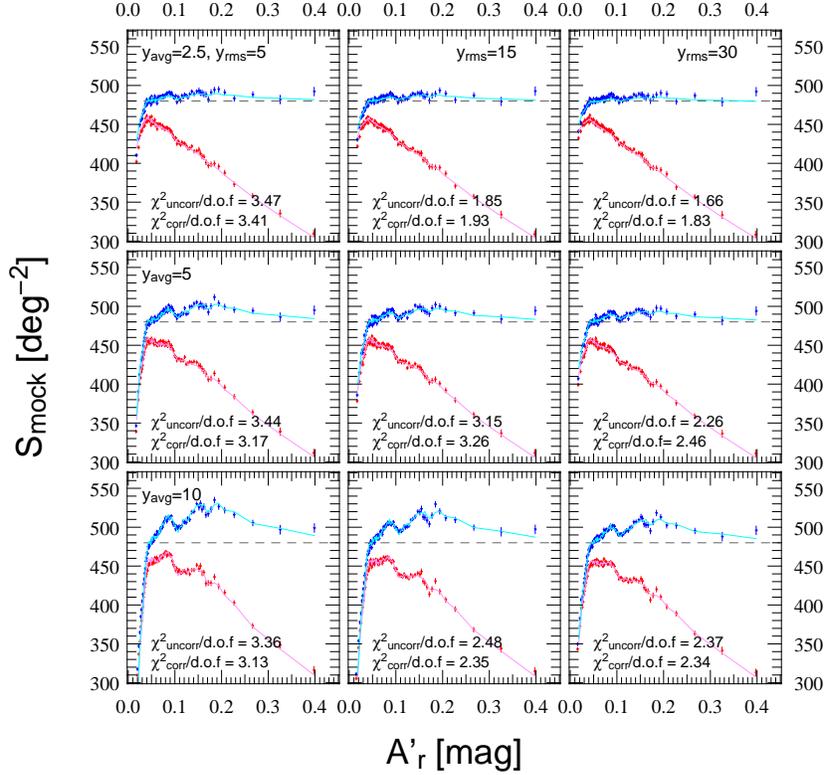}
  \end{center}
\figcaption{The results of the mock simulations with Poisson distributed
sample for various parameters of the log-normal PDF of $y$.  The symbols
indicate the results of the simulation for the mock Poisson sample, the
same as Figure \ref{fig:fiducial-parameters}.
The error bars reflect the Poisson noise alone.
The cyan and pink lines indicate the analytic model prediction from
equations (\ref{eq:surface-number-density-after}) and
(\ref{eq:surface-number-density-before}) in
\S \ref{sec:analytic}.  The lines and symbols are the same as
Figure \ref{fig:fiducial-parameters}.  The goodness of agreement between 
Poisson mock simulation and analytic model are evaluated by reduced $\chi^2$ for extinction 
un-corrected/corrected one, where only Poisson noise is considered. For all panels, the same
average surface number density, $\bar{S}=480 \mathrm{deg}^{-2}$, is
assumed and shown as gray dashed lines.  \label{fig:various-parameters}}
\end{figure*}

Finally we examine the dependence of the surface number densities on the
parameters of $y_{\rm avg}$ and $y_{\mathrm{rms}}$ for log-normal PDF of
$y$ (Fig. \ref{fig:various-parameters}).  The results indicate stronger
correlations for larger $y_{\rm avg}$, but turn out to be relatively
insensitive to $y_{\rm{rms}}$. This is why we choose $y_{\rm avg}$ and
$y_{\rm rms}$, instead of $\mu$ and $\sigma$, to parametrize the
log-normal PDF. A closer look reveals that larger $y_{\rm rms}$ 
shows slightly weaker anomaly, since a larger fraction of the mock galaxies 
are brighter than the IRAS/PSCz flux limit and does not contribute to FIR contamination.
This effect of flux limit becomes critical for very large $y_{\rm avg}$ and $y_{\rm rms}$, 
as we will see in \S \ref{subsec:fitting}.

As seen above, the mock result adopting equation
(\ref{eq:y-value-entire}) estimated for the {\it entire} SDSS galaxies
(Fig \ref{fig:fiducial-parameters}) indicates disagreement with the
observed anomaly (Fig \ref{fig:photometric-galaxy}).  This result may
appear to imply that the hypothesis of galaxy FIR contamination fails to
explain the observed anomaly. This is, however, not the case because we
have neglected spatial clustering of galaxies. The previous parameters
for the entire SDSS are estimated from the contribution of each single
galaxy itself, but in the presence of galaxy clustering, the FIR
emission associated with that galaxies can be significantly enhanced by
the neighbor galaxies.  In fact, the stacking analysis on the SFD map
revealed that the FIR emission of neighbor galaxies dominate the central
galaxy even by an order of magnitude \citep{Kashiwagi;2013}.  Therefore,
we should adopt $y_{\rm avg}$ and $y_{\rm rms}$ that represent the total
contribution both for each single galaxy and clustering neighbor
galaxies, in order to reproduce the observed anomaly by our Poisson mock
simulation.

  In principle, we can probe such FIR fluxes from the
comparison between mock simulations and observations, but the
simulations are very time-consuming. Thus we develop an analytic model
that reproduces the mock results in the next section.

%%%%%%%%%%%%%%%%%%%%%%%%%%%%%%%%%%%%%%%%%%%%%%%%%
%%%%%%%%%%%%%%%%%%%%%%%%%%%%%%%%%%%%%%%%%%%%%%%%%%

%%%%%%%%%%%%%%%%%%%%%%%%%%%%%%%%%%%%%%%%%%%%%%%%%%
\section{Analytic model of the FIR contamination}  
\label{sec:analytic}
%%%%%%%%%%%%%%%%%%%%%%%%%%%%%%%%%%%%%%%%%%%%%%%%%%

In this section, we develop an analytic model that describes the anomaly
of surface number densities of galaxies due to their FIR emission.  The
reliability of the analytic model is checked against the result of the
numerical simulations presented in the previous section.  We present a
brief outline in the next subsection, and the details are described in
Appendix \ref{app:analytic-detail}.

\subsection{Outline} \label{subsec:outline}

Let $A$ define the {\em true} Galactic extinction, not contaminated by
the galaxy emissions.  We denote the sky area whose value of the {\em
true} extinction is between $A$ and $A+dA$ by $\Omega (A)dA$, and the
number of galaxies that are located in the area $\Omega (A)dA$ by
$N_{\mathrm{gal}}(A)dA$.  Since there is no spatial correlation between
galaxies and the Galactic dust, the corresponding surface number
densities of the galaxies as a function of $A$:
%%%%%%%%%%%%%%%%%%%%%%%%%%%%%%%%%%%%%%
\begin{equation}
S(A) \equiv \frac{N_{\mathrm{gal}}(A)}{\Omega(A)}
\label{eq:intrinsic-distribution-function}
\end{equation}
%%%%%%%%%%%%%%%%%%%%%%%%%%%%%%%%%%%%%%
should be independent of $A$ and constant within the statistical
error.

If the FIR emission from galaxies contaminates the {\em true} extinction, 
however, the above quantities should depend on the 
contaminated extinction, $A^{\prime}$, which are defined as
$\Omega^{\prime}(A^{\prime})$ and
$N^{\prime}_{\mathrm{gal}}(A^{\prime})$, respectively.  Thus the {\it
observed} surface number densities, $S^{\prime}(A^{\prime})$, should be
%%%%%%%%%%%%%%%%%%%%%%%%%%%%%%%%%%%%%%%
\begin{equation}
S^{\prime}(A^{\prime}) 
= \frac{N_{\mathrm{gal}}^{\prime}(A^{\prime})}
{\Omega^{\prime}(A^{\prime})}.
\label{eq:contaminated-surface-number-density}
\end{equation}
%%%%%%%%%%%%%%%%%%%%%%%%%%%%%%%%%%%%%%%
The essence of our analytic model is how to compute the expected
$\Omega^{\prime}(A^{\prime})$ and
$N_{\mathrm{gal}}^{\prime}(A^{\prime})$ under the presence of the FIR
contamination of galaxies, which are distorted from the given {\em true}
$\Omega(A)$ and $N_{\rm{gal}}(A)$.

Due to its angular resolution, the FIR emission of multiple galaxies 
contaminate to the extinction in the SFD map at a given position.
Thus we need to sum up the FIR emission contribution of those galaxies 
located within the angular resolution scale:
%%%%%%%%%%%%%%%%%%%%%%%%%%%%%%%%%%%%%%%%
\begin{equation}
A^{\prime} = A + \Delta A, 
\label{eq:A-prime}
\end{equation}
%%%%%%%%%%%%%%%%%%%%%%%%%%%%%%%%%%%%%%%%
where the additional extinction, $\Delta A$, is computed
by summing up the contribution of the $i$-th galaxies ($i=1\sim N$) 
located in the pixel:
%%%%%%%%%%%%%%%%%%%%%%%%%%%%%%%%%%%%%%%%
\begin{equation}
\Delta A = \sum_{i=1}^N \Delta A_i .
\label{eq:DeltaA}
\end{equation}
%%%%%%%%%%%%%%%%%%%%%%%%%%%%%%%%%%%%%%%%

In order to perform the summation analytically, we need a joint
probability distribution function, $P_{\mathrm{joint}}(\Delta A,N)$,
corresponding to the situation where there are $N$ galaxies in a pixel 
of the dust map, and the total contribution of those galaxies is $\Delta A$. In Appendix
\ref{app:analytic-detail}, we present a prescription to compute
$P_{\mathrm{joint}}(\Delta A,N)$, and provide the integral
expressions for $\Omega^{\prime}(A^{\prime})$ and
$N_{\mathrm{gal}}^{\prime}(A^{\prime})$.

\subsection{Application of the analytic model}
\label{subsec:apply}

The analytic expressions for $\Omega^{\prime}(A^{\prime})$,
$N^{\prime}_{\rm{gal,corr}}(A^{\prime})$ and
$N^{\prime}_{\rm{gal,uncorr}}(A^{\prime})$ are given in equations
(\ref{eq:Omega^prime}), (\ref{eq:number-after}) and
(\ref{eq:number-before}) in Appendix \ref{app:analytic-detail}. Thus one
can compute the surface number densities for the $i$-th subregion of the
extinction between $A_i^{\prime}$ and $A_{i+1}^{\prime}$ as
%%%%%%%%%%%%%%%%%%%%%%%%%%%%%%%%%%%%%%%%
\begin{eqnarray}
S^{\prime}_{\rm{corr},i}
&=&
\frac{\int_{A_i^{\prime}}^{A_{i+1}^{\prime}} 
N^{\prime}_{\rm{gal,corr}}(A^{\prime})dA^{\prime}}
{\int^{A_{i+1}^{\prime}}_{A_i^{\prime}}
\Omega^{\prime}(A^{\prime})dA^{\prime}}, 
\label{eq:surface-number-density-after}\\
S^{\prime}_{\rm{uncorr},i}
&=&
\frac{\int_{A_i^{\prime}}^{A_{i+1}^{\prime}} 
N^{\prime}_{\rm{gal, uncorr}}(A^{\prime})dA^{\prime}}
{\int^{A_{i+1}^{\prime}}_{A_i^{\prime}}
\Omega^{\prime}(A^{\prime})dA^{\prime}},
\label{eq:surface-number-density-before}
\end{eqnarray}
%%%%%%%%%%%%%%%%%%%%%%%%%%%%%%%%%%%%%%%%
where $S^{\prime}_{\rm{corr}}$ and $S^{\prime}_{\rm{uncorr}}$ are the
extinction-corrected and uncorrected surface number densities,
respectively.  The solid lines in Figure
\ref{fig:various-parameters} show the surface number densities
calculated from equations (\ref{eq:surface-number-density-after}) and
(\ref{eq:surface-number-density-before}) adopting 9 parameter sets of $y_{\rm avg}$ 
and $y_{\rm rms}$.  The horizontal axis, an average
extinction in each subregion, is calculated as
%%%%%%%%%%%%%%%%%%%%%%%%%%%%%%%%%%%%%%%%
\begin{eqnarray}
A^{\prime}_{\rm{corr},i}
&=&
\frac{\int_{A_i^{\prime}}^{A_{i+1}^{\prime}} 
A^{\prime}N^{\prime}_{\rm{gal,corr}}(A^{\prime})dA^{\prime}}
{\int^{A_{i+1}^{\prime}}_{A_i^{\prime}}
N^{\prime}_{\rm{gal,corr}}(A^{\prime})dA^{\prime}},  \\
A^{\prime}_{\rm{uncorr},i}
&=&
\frac{\int_{A_i^{\prime}}^{A_{i+1}^{\prime}} 
A^{\prime}N^{\prime}_{\rm{gal,uncorr}}(A^{\prime})dA^{\prime}}
{\int^{A_{i+1}^{\prime}}_{A_i^{\prime}}
N^{\prime}_{\rm{gal,uncorr}}(A^{\prime})dA^{\prime}}.
\label{eq:average-extinction}
\end{eqnarray}
%%%%%%%%%%%%%%%%%%%%%%%%%%%%%%%%%%%%%%%%

Figure \ref{fig:various-parameters} clearly indicates that the analytic
predictions and the simulation results are in good agreement.  Strictly
speaking, the agreement is not perfect in a sense that the reduced
$\chi^2$ is as large as $\sim$ 3.5 for the worst cases, when only the
Poisson noise is considered.  The statistical errors for the observed
SDSS surface number densities (Figure \ref{fig:photometric-galaxy}),
however, includes the variance due to spatial clustering and are larger
than the Poisson noise by an order of magnitude. Thus the discrepancy
between the mock simulation and the analytic model is negligible for the
parameter-fit analysis to the observational result in the following
section.

\section{Comparison of FIR contamination with the observed anomaly}\label{sec:comparison}

Given the success of the analytic model described above, we
compare the model prediction with the observed SFD anomaly. Our
discussion in this section is organized as follows.

(1) We attempt to find the optimal values of $y_{\rm avg}$ and
$y_{\rm rms}$ by fitting the analytic model prediction to the observed
anomaly. It turns out that the observed anomaly is reproduced fairly
well with a relatively wide range of $y_{\rm avg}$ and $y_{\rm
rms}$ as long as $y_{\rm avg}$ is larger than $\sim 4$. 

(2) This value of $y_{\rm avg}$ should be compared with with the empirical, and thus
model-independent, result $y_{\rm avg} \approx 3.8$ obtained from the
stacking analysis \citep{Kashiwagi;2013}.  The fact that the rough
agreement of the two independent estimates for the average FIR to r-band
fluxes is interpreted as a supporting evidence for our FIR
explanation of the observed SFD anomaly.

(3) Finally, we attempt to reproduce the FIR flux of SDSS galaxies
required above within our framework of the simplified modeling for
FIR-to-optical relation. The estimated FIR flux qualitatively explains the result
(2), but not quantitatively.  We suspect that this is due to the
limitation of our FIR assignment model for galaxies, and not the basic
flaw of the FIR explanation for the SFD anomaly. Namely, given the fact
that the stacking analysis already indicates the barely required value
for $y_{\rm avg}$, we have to refine the FIR assignment model for SDSS
galaxies, rather than to rule out the FIR explanation itself.

\subsection{Estimating of the FIR emission of galaxies 
from the observed anomaly \label{subsec:fitting}}

Given the success of the analytic model described above, we attempt to 
find the best-fit parameters, $y_{\rm avg}$, and $y_{\mathrm{rms}}$, 
to the observed anomaly by minimizing 
%%%%%%%%%%%%%%%%%%%%%%%%%%%%%%%%%%%%%%%%%
\begin{equation}
\label{eq:chi-square}
\chi^2(y_{\rm avg},y_{\mathrm{rms}},\bar{N}) 
= \sum_i 
\frac{(S^{\rm{obs}}_{\rm{uncorr},i} 
-S^{\prime}_{\rm{uncorr},i})^2}{\sigma_{\rm{obs},i}^2},
\end{equation}
%%%%%%%%%%%%%%%%%%%%%%%%%%%%%%%%%%%%%%%%%
where $S^{\rm{obs}}_{\rm{uncorr},i}$ is the extinction-{\it uncorrected}
surface number densities in the $i$-th subregion of extinction,
$\sigma_{\mathrm{obs},i}$ is its statistical errors, and
$S^{\prime}_{\rm{uncorr},i}=S^{\prime}_{\rm{uncorr},i} (y_{\rm
avg},y_{\mathrm{rms}},\bar{N})$ is the analytic model prediction given
by equation (\ref{eq:surface-number-density-before}).  In the present
fit, we use the extinction-uncorrected surface number densities, but the
result is almost the same even if we use $S_{\rm{corr}}$ instead.  In
addition to $y_{\rm avg}$ and $y_{\rm rms}$, we include another free
parameter, the intrinsic average number of galaxy in a pixel, $\bar{N}$,
which is also unknown since the extinction correction is not necessarily
reliable. It turns out that $\bar{N}$ is in the range of $480$ to $500 \rm{[deg^{-2}]}$ 
and the results below is not sensitive to this value.

\begin{figure*}
 \begin{center}
  \includegraphics[width=0.348\textwidth]{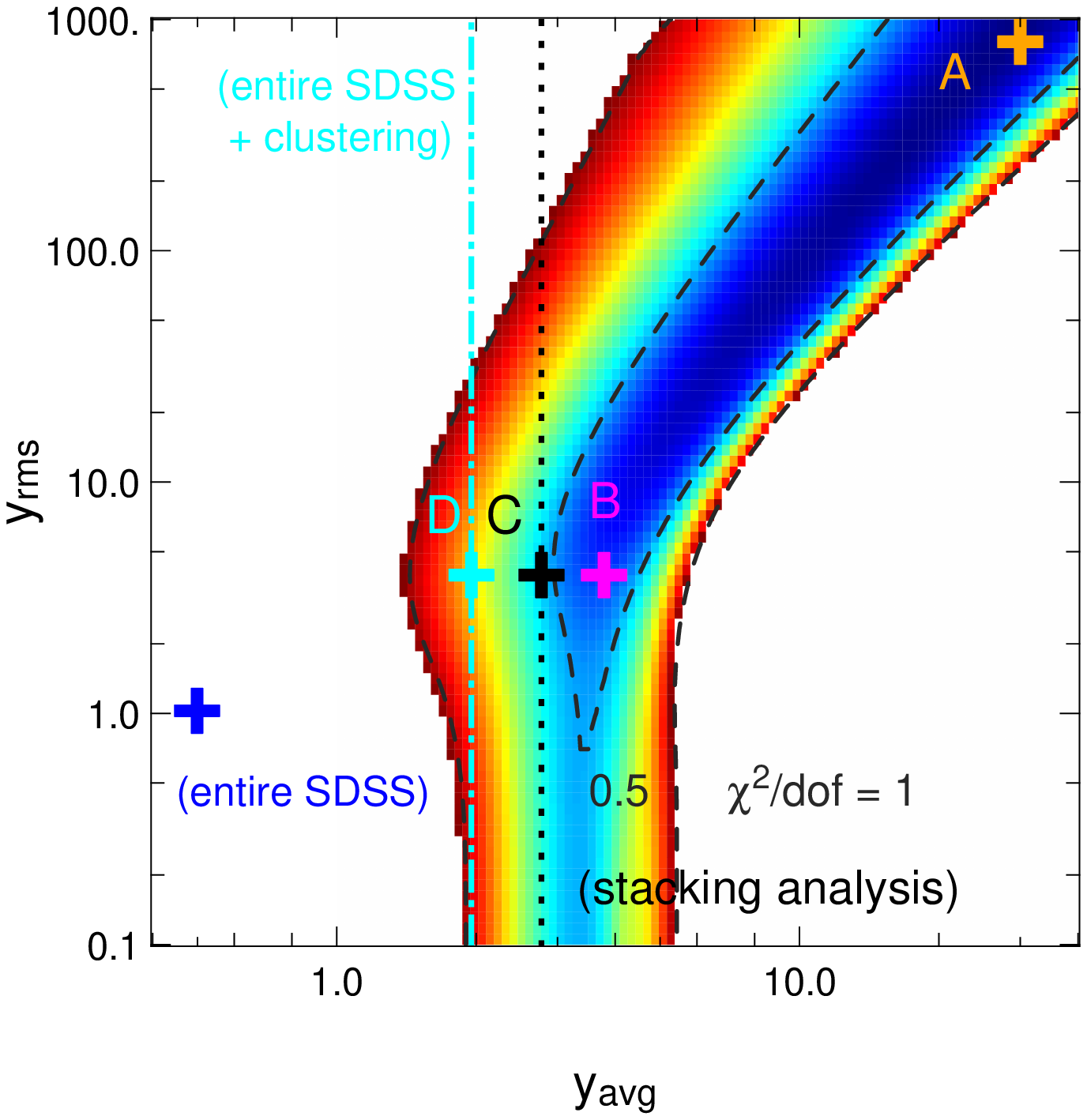}
  \hspace{5mm}
  \includegraphics[width=0.355\textwidth]{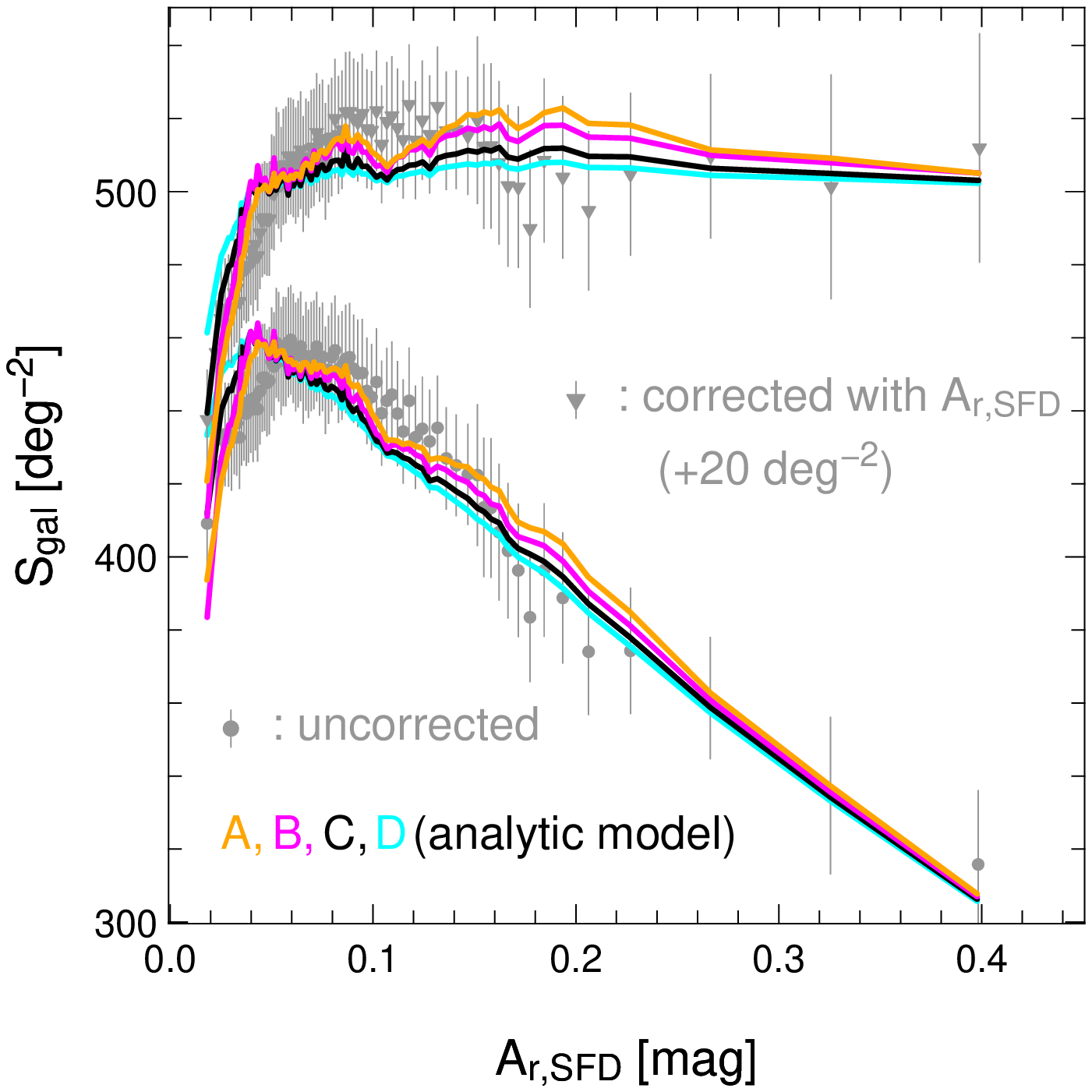}
  \bigskip
  
  \includegraphics[width=0.36\textwidth]{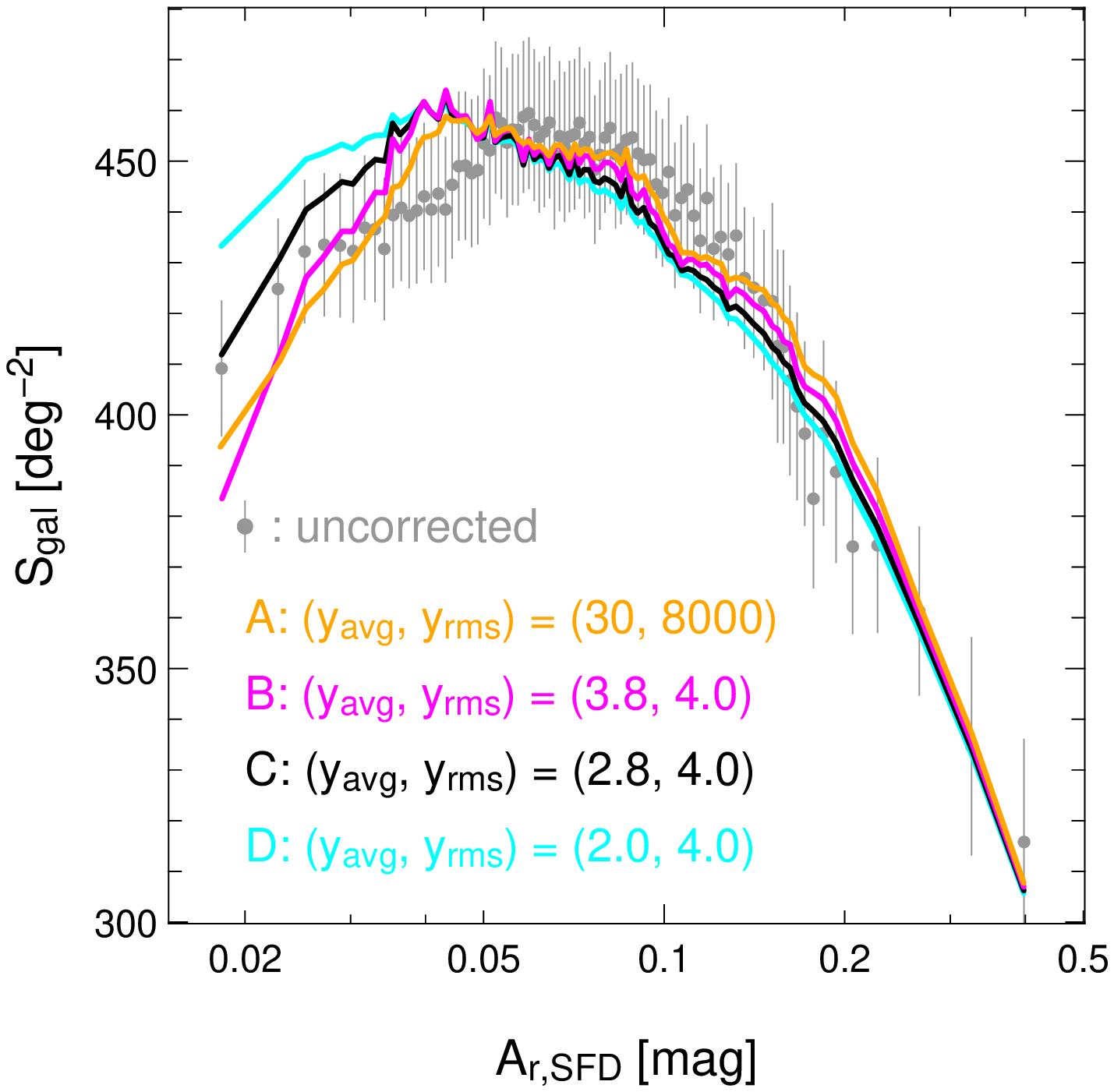}
  \hspace{2.5mm}
  \includegraphics[width=0.36\textwidth]{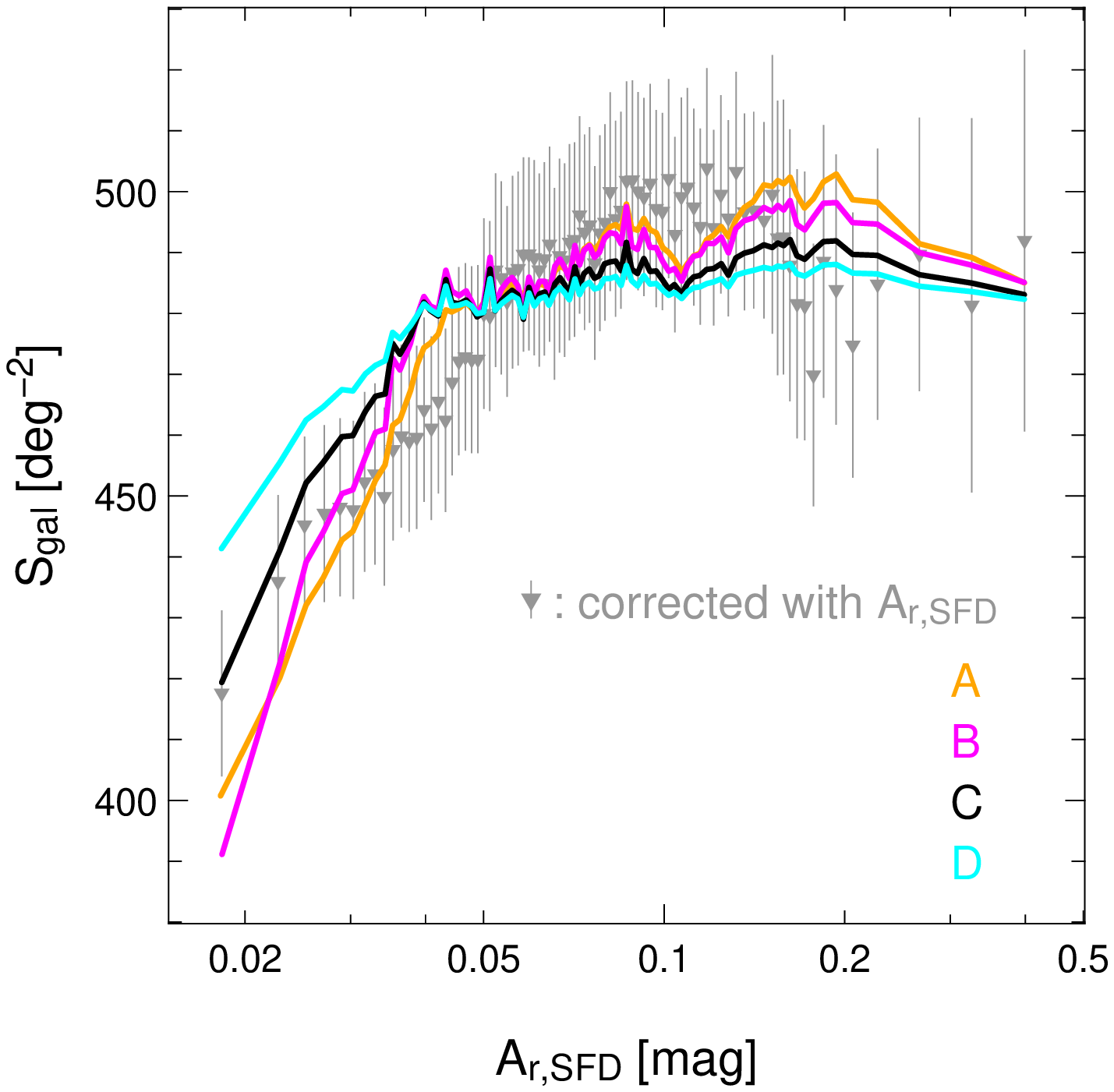}
  \end{center}
\figcaption{Fit to the observed anomaly using the analytical model.
{\em top left panel}; constraints on $y_{\rm avg}$ and
$y_{\mathrm{rms}}$ through the chi-squared analysis with equation
(\ref{eq:chi-square}).  The black dashed curves correspond to
$\chi^2/{\rm d.o.f}=1$ and $\chi^2/{\rm d.o.f}=0.5$ constraints. The
orange (A) and magenta (B) crosses are representative values that best
explain the observed anomaly.  The black dotted line and cross (C)
indicates the value of $y_{\rm avg}$ estimated by stacking analysis
\citep{Kashiwagi;2013}. The blue cross shows the best-fit parameters for
single galaxy of entire SDSS sample estimated in
\S\ref{subsec:IRAS_SDSS}. The cyan dot-dashed line and cross (D) also
indicates the value of $y_{\rm avg}$ estimated for entire SDSS sample,
but including neighbor galaxies contribution
(\S\ref{subsec:clustering}).  {\em top right panel}; the analytic model
predictions plotted over the observational data.  The solid lines
indicate the analytic prediction by equation
(\ref{eq:surface-number-density-after}) and
(\ref{eq:surface-number-density-before}), adopting the values of
$(y_{\rm avg}, y_{\rm rms})$ shown as the crosses in {\em top left}. The
symbols are the observational results for the SDSS galaxies in $r$-band,
the same as Figure \ref{fig:photometric-galaxy}.  The plots for $S_{\rm
gal}$ corrected with $A_{r,{\rm SFD}}$ are shifted by $+20{\rm
deg}^{-2}$ just for clarity.  {\em bottom left}; the same as {\em top
right}, but indicates $S_{\rm gal}$ uncorrected for extinction and the
horizontal axis is log-scaled.  {\em bottom left}; the same as {\em
bottom left}, but for $S_{\rm gal}$ corrected with $A_{r,{\rm SFD}}$.
\label{fig:figure10}}
\end{figure*}

In reality, however, the resulting constraints are not so strong as
shown in the top-left panel in Figure \ref{fig:figure10}. This is partly
due to the fact that we simply compute $\sigma_{\mathrm{obs},i}$ from
the variance of each extinction bin, which does not represent the proper
error. Thus our analysis here should be interpreted as a qualitative
attempt to find a possible parameter space to explain the anomaly in
terms of the FIR contamination; it would be quite difficult to make more
quantitative analysis, given several crude approximations in
our theoretical modeling and the poor angular-resolution and uncertain
dust temperature correction in the SFD map.

Bearing this remark in mind, let us consider the constraints on
$y_{\rm avg}$ -- $y_{\rm{rms}}$ plane from the observed anomaly shown in
the top-left panel of Figure \ref{fig:figure10}. Fairly acceptable fits
are obtained over the bluish region. Just for illustration, we select
two widely separated points A and B with $(y_{\rm avg}, y_{\rm
rms})=(30, 8000)$ and $(3.8, 4.0)$, respectively, and plot the
corresponding analytical predictions in the other three panels. Even
though their $y_{\rm avg}$ is different by an order of magnitude, the
two sets of parameters account for the observed anomaly reasonably and
equally well.

%%%%%%%%%%%%%%%%%%%%%%%%%%%%%%%%%%%%%%%%%%%%%%%
%%%%%%%%%%%%%%%%%%%%%%%%%%%%%%%%%%%%%%%%%%%%%%%%

\subsection{Comparison with the stacking image analysis}
\label{subsec:comparison-stacking}

We have shown that the anomaly in the surface number densities of SDSS
galaxies on the SFD map is well reproduced by assuming their 100$\mum$
to $r$-band flux ratio is $\sim 3.8$ on average, where the 100$\mum$
flux includes the contribution of neighbor galaxies. On the other hand,
the flux ratio of a single galaxy is estimated as $\sim 0.5$ (see \S
\ref{subsec:IRAS_SDSS}).

Indeed these values should be compared with the result of the stacking
image analysis by \citet{Kashiwagi;2013}. They stacked the SDSS galaxies
on the SFD map and found that a galaxy of $r$-band magnitude $m_r$
contributes to the extinction on average by
%%%%%%%%%%%%%%%%%%%%%%%%%%%%%%%%%%%%%%%%%%%%%%
\begin{equation}
\label{eq:dA-mr-gal-single}
\Delta A_r^{\rm s}(m_r) = 0.087 \times 10^{0.41(18-m_r)}~{\rm [m mag]},
\end{equation}
%%%%%%%%%%%%%%%%%%%%%%%%%%%%%%%%%%%%%%%%%%%%%%
by itself (single term), and 
%%%%%%%%%%%%%%%%%%%%%%%%%%%%%%%%%%%%%%%%%%%%%%
\begin{equation}
\label{eq:dA-mr-gal-total}
\Delta A_r^{\rm tot}(m_r) = 0.64 \times 10^{0.17(18-m_r)}~{\rm [m mag]},
\end{equation}
%%%%%%%%%%%%%%%%%%%%%%%%%%%%%%%%%%%%%%%%%%%%%%
including the contribution from neighbor galaxies, corresponding to the
clustering term in \citet{Kashiwagi;2013}. The above extinction due to
the $100\mum$ emission from galaxies is translated into its $100\mum$ to
$r$-band flux ratio as
%%%%%%%%%%%%%%%%%%%%%%%%%%%%%%%%%%%%%%%%%%%%%%%%
\begin{equation}
\label{eq:dA-y-avg}
y= \frac{2\pi \sigma^2}{f_r \nu_r / \nu_{\rm 100\mum}} 
\frac{\Delta A_r}{k_r p},
\end{equation}
%%%%%%%%%%%%%%%%%%%%%%%%%%%%%%%%%%%%%%%%%%%%%%%%
where $\sigma$ is the Gaussian PSF width and $f_r$ is the $r$-band flux.
Thus integrated over the differential number density, equations
(\ref{eq:dA-mr-gal-single}) and (\ref{eq:dA-mr-gal-total}) suggest that
%%%%%%%%%%%%%%%%%%%%%%%%%%%%%%%%%%%%%%%%%%%%%%%%
\begin{eqnarray}
\label{eq:y-avg-single}
\bar{y}_{\rm avg}^{\rm s} = 
\frac{\int dm_r\frac{dN}{dm_r}y_{\rm avg}^{\rm s}(m_r)}
{\int dm_r\frac{dN}{dm_r}} = 0.239,
\end{eqnarray}
%%%%%%%%%%%%%%%%%%%%%%%%%%%%%%%%%%%%%%%%%%%%%%%%
and
%%%%%%%%%%%%%%%%%%%%%%%%%%%%%%%%%%%%%%%%%%%%%%%%
\begin{eqnarray}
\label{eq:y-avg-clustering}
\bar{y}_{\rm avg}^{\rm tot} = 
\frac{\int dm_r ({dN}/{dm_r}) y_{\rm avg}^{\rm c}(m_r)}
{\int dm_r ({dN}/{dm_r})} = 2.77,
\end{eqnarray}
%%%%%%%%%%%%%%%%%%%%%%%%%%%%%%%%%%%%%%%%%%%%%%%%
respectively.

These values are based on the direct measurement of the FIR
contamination, and thus independent of the modeling of 100$\mum$ to
optical relation. We also emphasis that they should automatically
include possible contributions from those galaxies not identified by
SDSS.  Therefore the sum of the two terms can be reliably interpreted as
the expected contribution of the SDSS galaxies to $y_{\rm avg}$
including neighbor galaxies, which is plotted in Figure
\ref{fig:figure10}. While we do not know the corresponding $y_{\rm
rms}$, we have already found that the dependence of the anomaly on
$y_{\rm rms}$ is rather weak, at least in our analytic model. Thus the
empirical value of $y_{\rm avg}$ from the stacking analysis roughly
explains the observed anomaly as plotted in the three panels of Figure
\ref{fig:figure10}.

We interpret this agreement as a supporting evidence for the FIR model of the SFD
anomaly given the fact that we assume a very simple relation between
100$\mum$ and optical luminosities, neglecting the galaxy morphology
dependence that certainly leads to the FIR flux difference.

\subsection{Estimates of clustering contribution of SDSS galaxies}
\label{subsec:clustering}

We tried to independently estimate $y_{\rm avg}$, including an
additional contribution of neighbor galaxies, using the SDSS galaxy
distribution over the SFD map, instead of the stacking result by
\citet{Kashiwagi;2013} discussed in \S \ref{subsec:comparison-stacking}.
We first randomly assign the FIR flux of SDSS galaxies assuming $(y_{\rm
avg}, y_{\rm rms})=(0.5, 1.0)$ {\it for each SDSS galaxy itself
neglecting the clustering term}.  Second, we sum up the FIR fluxes of
galaxies convolved with the PSF of the SFD map (the Gaussian width of
$3'.1$) centered at each galaxy. Finally we compute $y_{\rm avg}$ and
$y_{\rm rms}$ using the summed FIR fluxes after subtracting the average
background flux.

Note that the resulting values of $y_{\rm avg}$ and $y_{\rm rms}$ should
be diffrent from the above input values because of the contribution of
the clustering term.  We find $y_{\rm avg} \approx 2$, but $y_{\rm rms}$
is not well determined because it turned out to be very sensitive to the
choice of the background flux. This result indicates that the FIR flux
of the SDSS galaxies explains only a half of those required to well
reproduce the observed anomaly, $y_{\rm avg}=3.8$.

Indeed, employing $y_{\rm avg} \approx 2$, our model still reproduces
the anomaly qualitatively, but the predicted feature is substantially
weaker than that of the observed one.  The assigned FIR flux in this
model, however, is based on the single galaxy contribution estimated in
\S \ref{subsec:IRAS_SDSS} ($y_{\rm avg}=0.5$), thus would be sensitive
to the FIR assignment model. Given the fact that the empirical value
from the stacking analysis, which is independent of such models, is
fairly successful in reproducing the anomaly, we suspect that the factor
of two difference originates from the limitation of our crude modeling
for FIR flux, instead of the basic flaw of the FIR explanation of the
anomaly.

\section{Discussion}\label{sec:discussion}

\subsection{Effects of spatial clustering of galaxies}
\label{subsec:N-body}

Both the mock simulations and the analytic model discussed in the
previous section completely ignore the spatial clustering of
galaxies. We, therefore, examine the clustering effect on the anomaly in
this subsection. The most straightforward method is to replace the
Poisson distributed mock galaxies by dark matter particles from cosmological
N-body simulation.  For that purpose, we use a realization in the
standard $\Lambda\rm{CDM}$ cosmology with $\sigma_8=0.76$ performed by
\citet{Nishimichi;2009}.

We repeat similar mock observations as discussed in \S
\ref{subsec:poisson}, except for that we assign $r$-band luminosity to
each mock galaxy instead of their apparent magnitude.  To be specific,
(i) we randomly assign $r$-band luminosities to all N-body dark matter
particles according to the luminosity function of equation
(\ref{eq:LFopt}), (ii) convert their luminosities to apparent $r$-band
magnitudes observed from a fixed observer position, and (iii) randomly
select a fraction of the mock galaxies to match with the SDSS observed
$dN/dm_r$ (Figure \ref{fig:magnitude-distribution}).

We repeat the same fitting analysis as Figure \ref{fig:figure10},
except that the data are now replaced by the mock result on the basis of
the cosmological N-body simulation with $y_{\rm avg}=3.8$ and $y_{\rm
rms}=4.75$.  The mock observation including the galaxy clustering effect
result shows stronger anomaly than Poisson mock simulation with the
identical $y_{\rm avg}$ and $y_{\rm rms}$.  The analytic model that
neglects the spatial clustering still reproduces the simulated anomaly
very well, but the best-fit $y_{\rm avg}$ overestimate the real values
employed in the simulation by a factor of $\sim$2. Thus the clustering
effect can be absorbed effectively by re-interpreting the best-fit
values of $y_{\rm avg}$ appropriately. The clustering effect estimated here 
is largely consistent with the clustering term contribution estimated directly 
from the SDSS galaxies (\S \ref{subsec:clustering}).

In order to quantitatively understand the relation between this bias and
the strength of the galaxy spatial clustering, we have to incorporate
the effect of spatial clustering in our analytic model.  For that
purpose, we measure the PDF of the number of the N-body mock particles
in a pixel and replace the Poisson distribution in equation
(\ref{eq:P_joint}) with the measured one. The analytic model prediction,
however, hardly changes by such a modification.  Thus more sophisticated
improvements seem to be needed to account for the spatial clustering
effect, which is beyond the scope of this paper.

\subsection{Limitation of the correction for the FIR emission of galaxies}

\begin{figure*}
  \begin{center}
  \includegraphics[width=0.35\textwidth]{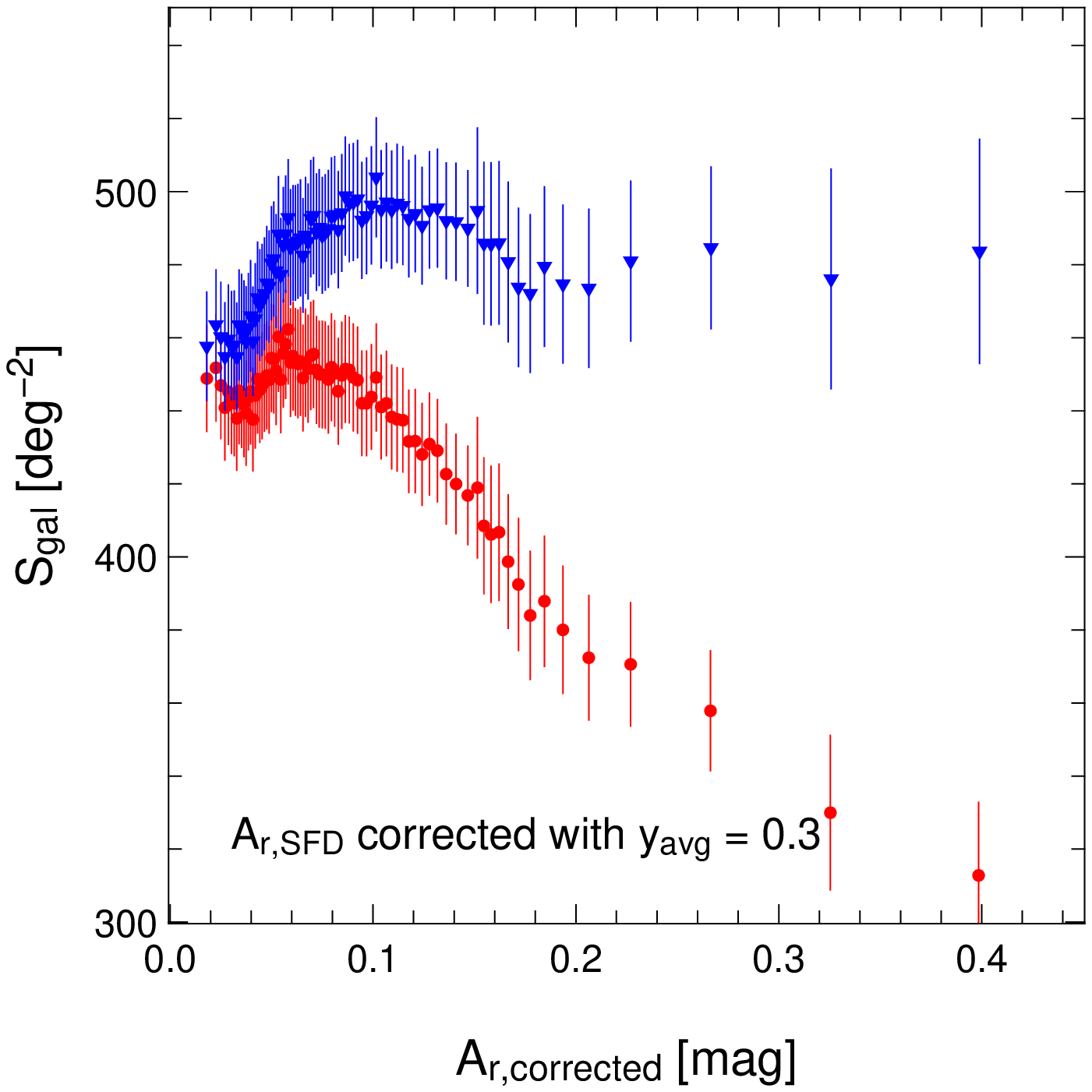}
  \hspace{20pt}
  \includegraphics[width=0.35\textwidth]{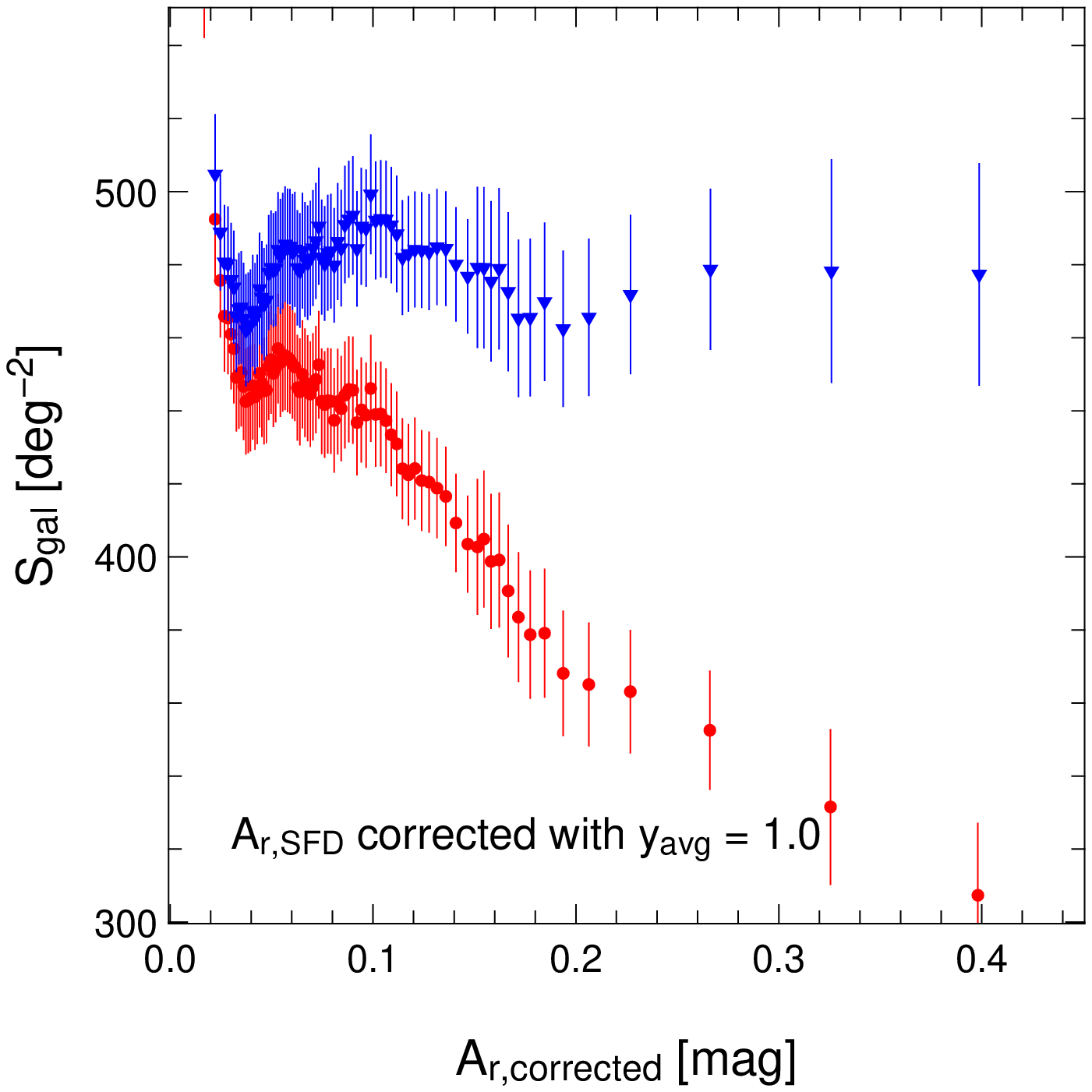}
  \bigskip
  
\includegraphics[width=0.35\textwidth]{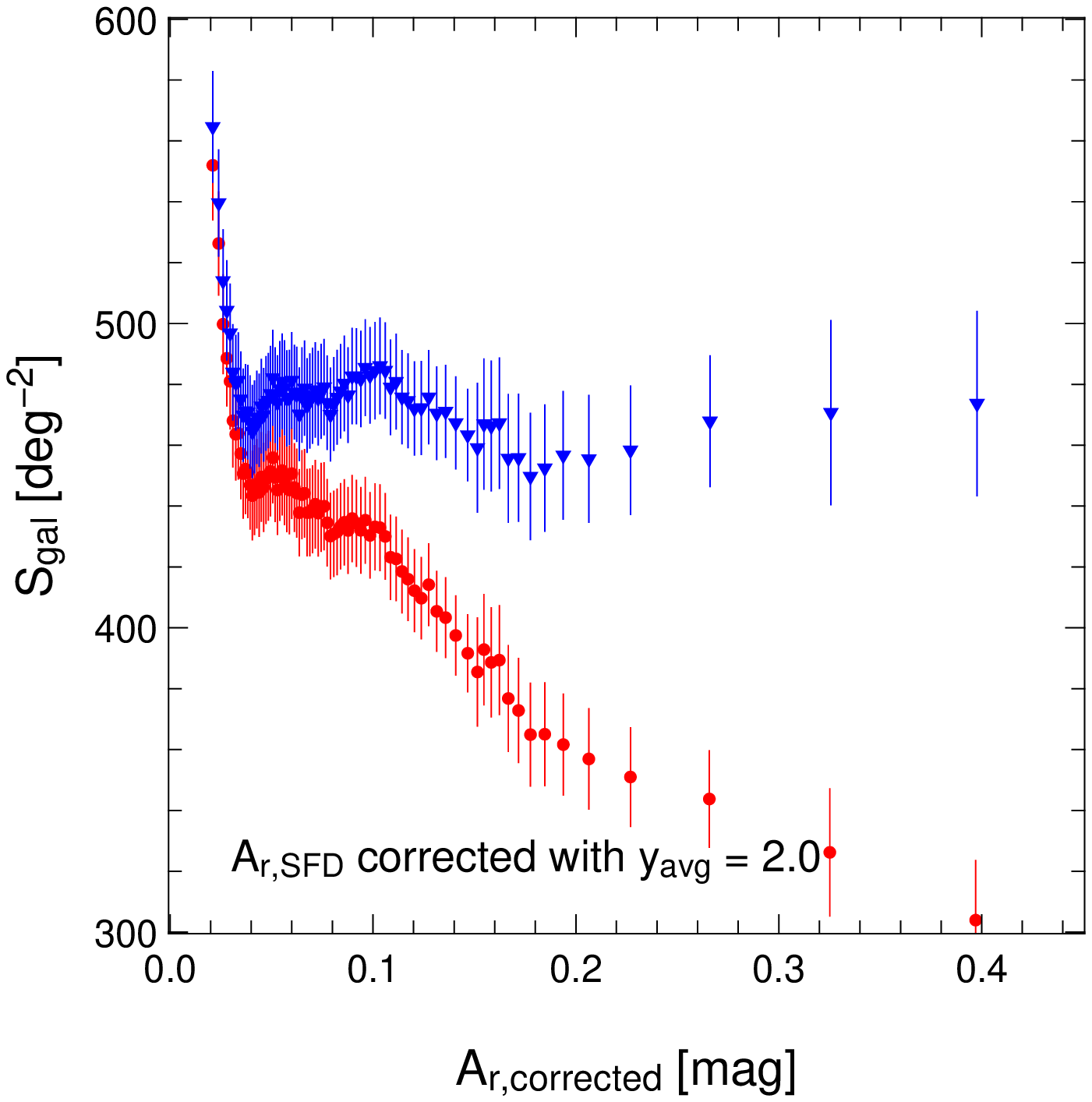}
\hspace{20pt}
  \includegraphics[width=0.35\textwidth]{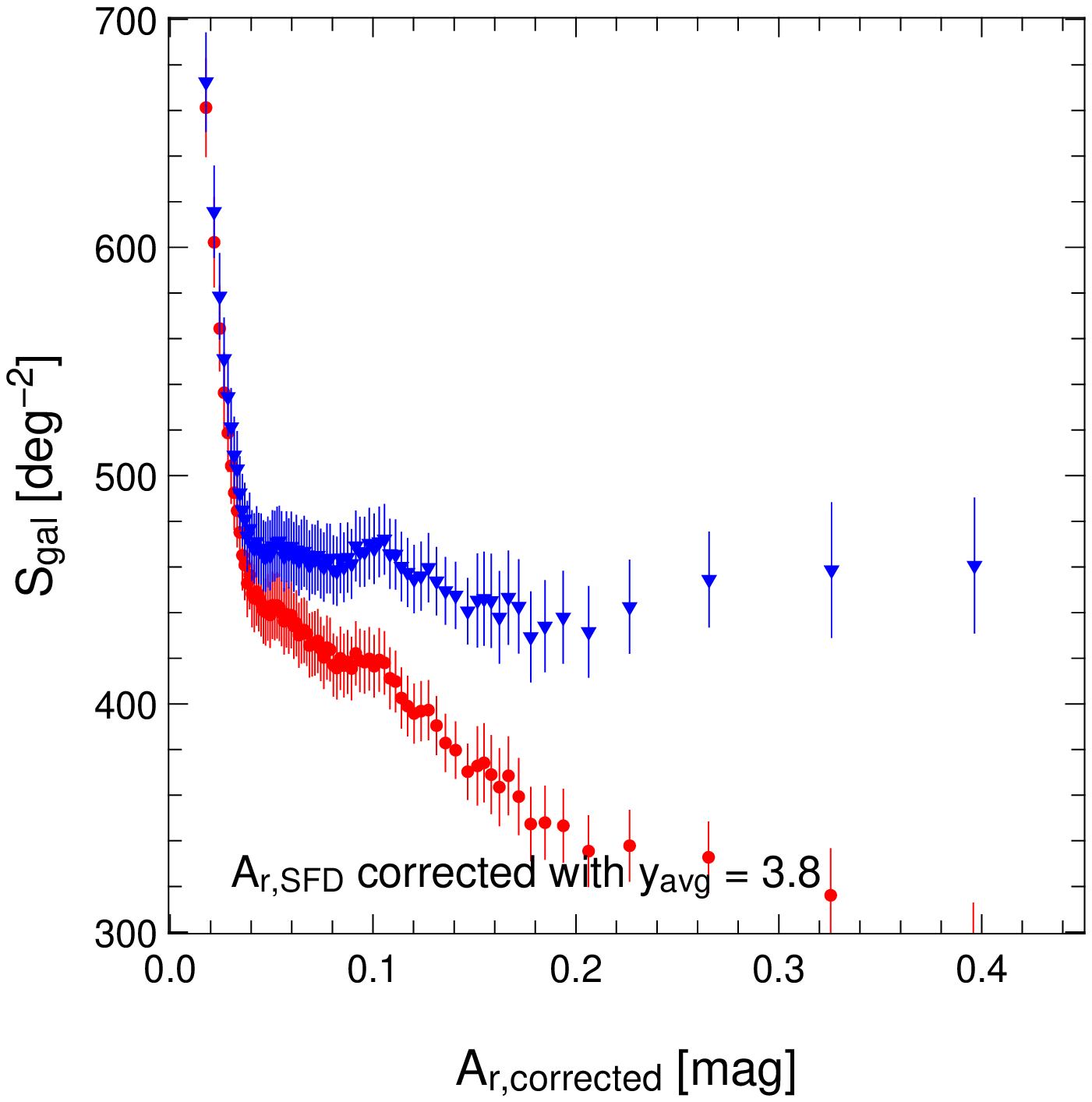}
  \end{center}
\figcaption{Surface number densities of the SDSS galaxies with $17.5<m_{\rm
 r}<19.4$ after subtracting their average FIR emission contamination, where
$y_{\rm avg} = 0.3, 1.0, 2.0, 3.8$ are adopted for estimation of the FIR emission of the SDSS galaxies.
\label{fig:sgal-after-contamination-correction}}
\end{figure*}

We attempt to correct the SFD map by subtracting the average FIR
contamination of SDSS galaxies. The corrected extinction at an angular
position $\mathbf\theta$ in the Galactic map is computed as
%%%%%%%%%%%%%%%%%%%%%%%%%%%%%%%%%%%%%%%%%%%%%%%%%%%%%%%%%%%%%%%%%%%%
\begin{equation}
\label{eq:dA-mr-pix}
A_{r,{\rm corrected}}({\mathbf\theta}) 
= A_{r,\rm SFD}({\mathbf\theta}) 
- \sum_{j} \Delta A({\mathbf\theta_j - \mathbf\theta};m_r^j),
\end{equation}
%%%%%%%%%%%%%%%%%%%%%%%%%%%%%%%%%%%%%%%%%%%%%%%%%%%%%%%%%%%%%%%%%%%%
where ${\mathbf\theta_j}$ is the position of the $j$-th galaxy with its
$r$-band magnitude of $m_r^j$.  We employ 4 different values for $\Delta
A$ given the uncertainty of the interpretation of the best-fit value of
$y_{\rm avg}$ discussed before. As shown in Figure
\ref{fig:sgal-after-contamination-correction}, however, the above
correction does not seem to remove the anomaly so well.  This results
may imply that the dependence of FIR properties on galaxy population,
which is neglected in our modeling, is essential for accurate correction
for the FIR contamination. As a future work, such a morphology
dependence of FIR luminosities of SDSS galaxies will be investigated by
stacking analysis, especially using recent high resolution diffuse FIR
measurements by AKARI \citep{Murakami;2007}, WISE \citep{Wright;2010},
etc.

%%%%%%%%%%%%%%%%%%%%%%%%%%%%%%%%%%%%%%%%%%%%%%%%%
%%%%%%%%%%%%%%%%%%%%%%%%%%%%%%%%%%%%%%%%%%%%%%%%%%

\subsection{Testing the Peek and Graves correction map}
\label{subsec:PG}

In \S \ref{subsec:comparison-stacking}, we found that the observed
anomaly of the SDSS galaxies is roughly explained by the contamination
of galaxy FIR emission.  Nevertheless, the observed and predicted
surface number densities (Fig \ref{fig:figure10}) do not match
perfectly, which might be attributed to other possible systematics in
the SFD map.

In order to check the possible systematic effect, we use the
improved extinction map by \citet[hereafter PG]{Peek;2010}. They found
that the SFD map {\it under-predicts} extinction up to $\sim 0.1$ mag in
$r$-band, using the passively evolving galaxies as standard color
indicators.  Their method is complementary to our galaxy number count
analysis in a sense that they directly measure the reddening by the
Galactic dust.  Since the resolution of the PG correction map to SFD is
$4^{\circ}.5$, the FIR fluctuations due to the emission of galaxies are
not expected to be removed.  The PG correction map, however, may have
removed other systematics than the FIR contamination, which are not
considered in our analytic model at all.

To see if their correction affect the number count analysis and the
anomaly in the original SFD map, we repeat the same analysis described
in \S 6 using the PG map.  Basically, we find a very similar correlation
between $S_{\rm gal}$ and $A_{r,\rm{PG}}$, suggesting that the PG map
still suffers from the FIR contamination of galaxies as expected.  We
note, however, that our analytic model prediction exhibits slightly
better agreement for the PG map than for the SFD map.  This may
indicates that possible systematic errors in the SFD map other
than the FIR contamination is at least partially removed in the PG map.

\section{Summary and conclusions}
\label{sec:conclusions}

In the present paper, we have revisited the origin of the anomaly of
surface number density of SDSS galaxies with respect to the Galactic
extinction, originally pointed out by \citet{Yahata;2007}.  We first
computed the anomaly using the SDSS DR7 photometric catalogs, and then
developed both numerical and analytic models to explain the anomaly.  We
take account of the contamination of galaxies in the IRAS $100{\mum}$
flux that was assumed to come entirely from the Galactic dust.  

Our main findings are summarized as follows.
%%%%%%%%%%%%%%%%%%%%%%%%%%%
\begin{itemize}
 \item Both numerical simulations and analytic model reproduce the
observed anomaly quite well. Thus we quantitatively confirmed the
validity of the hypothesis that the observed anomaly in the SFD Galactic
extinction map is mainly due to the FIR emission from galaxies,
originally proposed by \citet{Yahata;2007}.
\item  The comparison of the analytic model and the observed
anomaly constrains mainly the average $100\mum$ to optical
flux ratio for SDSS galaxies. The resulting value is in a reasonable
       agreement with 
that obtained from the stacking image analysis of the SDSS galaxies
by \citet{Kashiwagi;2013}.
\item We also independently estimated the FIR contribution of 
single SDSS galaxy based on IRAS/SDSS overlapped catalogue data 
assuming a simple relation between FIR and optical luminosities. Summing up 
such FIR flux according to the SDSS galaxy distribution, however, we find that 
those contribution only explains roughly half of that required to reproduce the 
observed anomaly. This result may be due to the limitation of our modeling of
the FIR to optical relation.
\end{itemize}
%%%%%%%%%%%%%%%%%%%%%%%%%%%

While our current analytic model still needs to be improved, the
fact that the empirically determined value of $y_{\rm avg}$ nicely
reproduces the observed anomaly indicates that the FIR emission of SDSS
galaxies is the major origin of the anomaly.

In particular, we note that subtracting the average FIR
contamination of the SDSS galaxies from the SFD extinction map does not
properly remove the observed anomaly.  This may imply that it is
essential to consider the dependence of FIR emission on galaxy
morphology and/or the effect of galaxy clustering, both of which we have
neglected in the current analytical model. Since morphology and spatial
clustering of galaxies are correlated in a complicated fashion, it is
not easy to identify the good strategy of the correction method. We are
currently working along this direction with the AKARI all-sky map data
in $60, 90, 140, 160\mum$.  The stacking image analysis of SDSS galaxies
with the higher-angular resolution map in multi-frequency bands would
enable us to estimate the FIR emission of galaxies as a function of
their properties including their color and morphology (T.Okabe et
al. 2015, in preparation).

The FIR contamination that explains the anomalous behavior in the
surface number density of the SDSS galaxies is just statistical and
tiny, on the order of (0.1$\sim$1)mmag of extinction in $r$-band, which
is much less serious than naively expected from the
anomaly. Nevertheless the galaxy FIR emission is correlated with the
large scale structure of the universe. Thus it may systematically bias
the cosmological analysis.  The present methodology is in principle
applicable to check the reliability, and even to improve the accuracy of
the future Galactic extinction map that should play a key role in all
astronomical observations, in particular for the purpose of precision
cosmology.

%%%%%%%%%%%%%%%%%%%%%%%%%%%%%%%%%%%%%%%%%%%%%%
%   Acknowledgements
%%%%%%%%%%%%%%%%%%%%%%%%%%%%%%%%%%%%%%%%%%%%%%
\acknowledgements

We thank Brice M{\'e}nard, Tsunehito Kohyama, Yasunori Hibi, and Hiroshi
Shibai for useful discussions. T.K and Y.S are grateful to the
hospitality of Department of Astrophysical Sciences, Princeton
University, where most of the present work was performed.
We also thank an anonymous referee for several constructive comments
and in particular for suggesting to compute the expected FIR fluxes
using the SDSS galaxy distribution as discussed in \S 6.3.

T.K. is supported by a Global COE Program "the Physical Sciences
Frontier", MEXT, Japan.  T.N. is supported by a Grant-in-Aid for the
JSPS fellows.  Y.S. gratefully acknowledges the supports from the Global
Collaborative Research Fund ``Worldwide Investigation of Other Worlds''
grant, the Global Scholars Program of Princeton University, and the
Grant-in Aid for Scientific Research by JSPS (No. 24340035).
A.T. acknowledges the support from Grant-in-Aid for Scientific Research
by JSPS (No. 24540257).

  Funding for the SDSS and SDSS-II has been provided by the Alfred
P. Sloan Foundation, the Participating Institutions, the National
Science Foundation, the U.S. Department of Energy, the National
Aeronautics and Space Administration, the Japanese Monbukagakusho, the
Max Planck Society, and the Higher Education Funding Council for
England.  The SDSS Web Site is http://www.sdss.org/.  The SDSS is
managed by the Astrophysical Research Consortium for the Participating
Institutions.  The Participating Institutions are the American Museum of
Natural History, Astrophysical Institute Potsdam, University of Basel,
University of Cambridge, Case Western Reserve University, University of
Chicago, Drexel University, Fermilab, the Institute for Advanced Study,
the Japan Participation Group, Johns Hopkins University, the Joint
Institute for Nuclear Astrophysics, the Kavli Institute for Particle
Astrophysics and Cosmology, the Korean Scientist Group, the Chinese
Academy of Sciences (LAMOST), Los Alamos National Laboratory, the
Max-Planck-Institute for Astronomy (MPIA), the Max-Planck-Institute for
Astrophysics (MPA), New Mexico State University, Ohio State University,
University of Pittsburgh, University of Portsmouth, Princeton
University, the United States Naval Observatory, and the University of
Washington.

\appendix
%%%%%%%%%%%%%%%%%%%%%%%%%%%%%%%%%%%%%%%%%%%%%%%%%%%%%%

%%%%%%%%%%%%%%%%%%%%%%%%%%%%%%%%%%%%%%%%
\section{Point Spread Function}
\label{app:psf}

In the mock simulation (\S \ref{sec:simulation}), we assign the FIR {\it
fluxes} to the mock galaxies by modeling the PDF of FIR to optical
luminosity ratio, $y$. On the other hand, their contribution to the
contamination in the SFD map is determined by their {\it intensities} as
%%%%%%%%%%%%%%%%%%%%%%%%%%%%%%%%%%%%%%%%
\begin{equation}
\Delta A_r = pk_r\frac{f_{100\mum}}{2\pi \sigma^2_{\rm eff}},
\end{equation}
%%%%%%%%%%%%%%%%%%%%%%%%%%%%%%%%%%%%%%%%
where $\sigma_{\rm eff}$ is the Gaussian width of the effective PSF,
thus the impact of the FIR contamination directly depends on
$\sigma_{\rm eff}$ even for the mock galaxies with the same $100\mum$
fluxes, $f_{100\mum}$.  Due to the smoothing effects by the pixelization
and interpolation of the SFD map, the effective PSF is degraded from
that applied in the mock simulations (${\rm FWHM}=5'.2$), which is aimed
to mimic the purely instrumental PSF.  Therefore, in order to precisely
reproduce the mock simulation results by our analytic model (\S
\ref{sec:analytic}), we have to carefully evaluate the appropriate
$\sigma_{\rm eff}$ to be applied in equation (\ref{eq:P_1}).  In this
appendix, we derive $\sigma_{\rm eff}$ as a function of the intrinsic
PSF width, $\sigma_{\rm int}$.

\begin{figure}
  \begin{center}
  \includegraphics[height=0.36\textwidth]{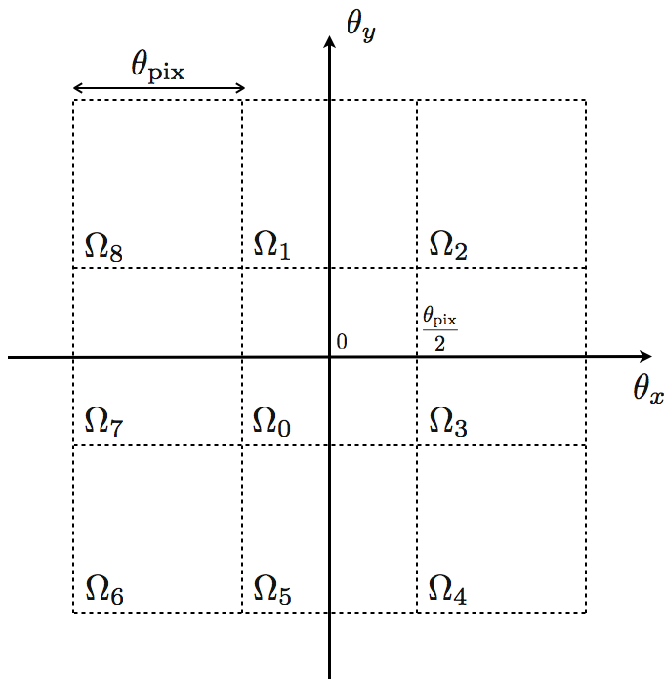}
    \includegraphics[height=0.36\textwidth]{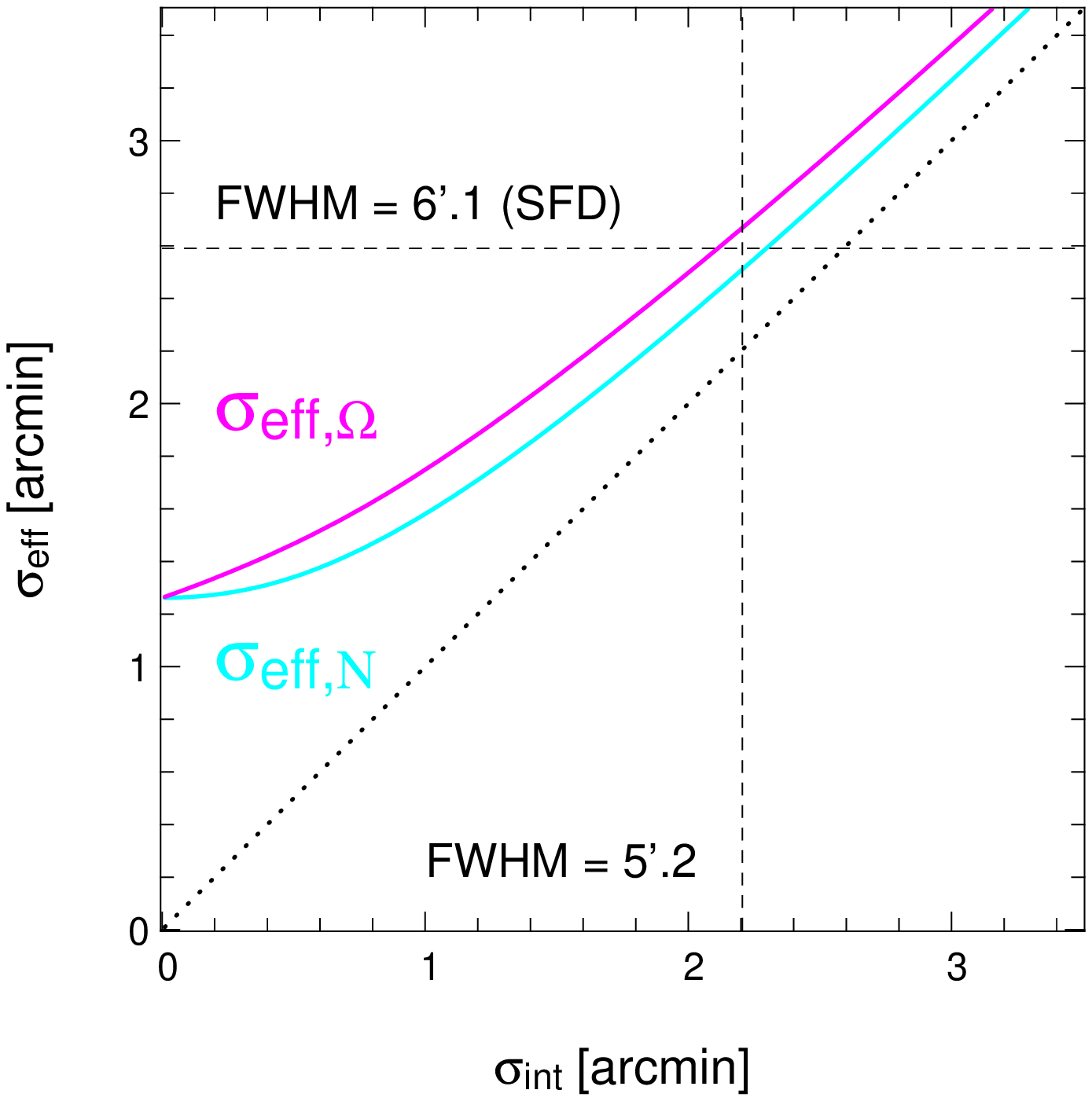}
  \end{center}
\figcaption{ {\em left panel;} Configuration of the SFD map pixels for calculating the effective PSF width. 
{\em right panel;} The effective Gaussian PSF widths, $\sigma_{{\rm eff},\Omega}$ (magenta) and $\sigma_{{\rm eff},N}$ 
(cyan), as functions of the intrinsic PSF width $\sigma_{\rm int}$. The vertical and horizontal dashed lines indicate 
the Gaussian PSF width applied in the mock simulation (\S \ref{sec:simulation}), and the resolution of the SFD map, respectively.
 \label{fig:sigma-eff}}
\end{figure}

First we calculate the intensity of a single galaxy with a given
$100\mum$ flux and position, taking into account of the two smoothing
effects.  Hereafter, we assume that the pixels of the SFD map are
squares with the sides, $\theta_{\rm pix}=2'.372$.  We denote the pixel
of the SFD map, in which the galaxy is located, as $\Omega_0$, and its
neighbor pixels as $\Omega_1$ to $\Omega_8$.  We define the
2-dimensional Cartesian coordinate system
$\mathbf{\theta}=(\theta_x,\theta_y)$, whose origin is at the center of
$\Omega_0$. The configuration of $\Omega_0$ to $\Omega_8$ is illustrated
in the left panel of Figure \ref{fig:sigma-eff}. The intensity of the
galaxy with $100\mum$ flux, $f$, in the pixel $\Omega_i$ ($i=0,...,8$)
is given as
%%%%%%%%%%%%%%%%%%%%%%%%%%%%%%%
\begin{equation}
I_{i}(\mathbf{\theta}_{\rm g}) = \frac{f}{2\pi \sigma_{\rm int}^2\Omega_{\rm pix}}
\int_{\Omega_i} \exp \left( -\frac{|\mathbf{\theta}-\mathbf{\theta}_{\rm g}|^2}{2\sigma_{\rm int}^2}\right) d\mathbf{\theta},
\end{equation}
%%%%%%%%%%%%%%%%%%%%%%%%%%%%%%%
where $\mathbf{\theta}_{\rm g}$ denotes the position of the galaxy, and
$\Omega_{\rm pix}=\theta_{\rm pix}^2$ is the area of the pixels. Since
the value of the SFD map extinction is evaluated by the linear CIC
interpolation, the intensity of the galaxy depends on $\theta_{\rm g}$,
but also the position where the value is evaluated, $\theta$, and
calculated as
%%%%%%%%%%%%%%%%%%%%%%%%%%%%%%%
\begin{eqnarray}
I_{\rm{CIC}}(\mathbf{\theta},\mathbf{\theta}_{\rm g})
=
\left(1-\frac{\theta_x}{\theta_{\rm{pix}}}\right)  \left(1-\frac{\theta_y}{\theta_{\rm{pix}}}\right) 
&I&_{i_1}(\mathbf{\theta}_{\rm g})
+
\left(1-\frac{\theta_x}{\theta_{\rm{pix}}}\right)  \frac{\theta_y}{\theta_{\rm{pix}}}
I_{i_2}(\mathbf{\theta}_{\rm g}) \cr
+
\frac{\theta_x}{\theta_{\rm{pix}}} \frac{\theta_y}{\theta_{\rm{pix}}}
&I&_{i_3}(\mathbf{\theta}_{\rm g})
+\frac{\theta_x}{\theta_{\rm{pix}}} \left(1-\frac{\theta_y}{\theta_{\rm{pix}}}\right) 
I_{i_4}(\mathbf{\theta}_{\rm g}),
\end{eqnarray}
%%%%%%%%%%%%%%%%%%%%%%%%%%%%%%%%
where $(i_1,...,i_4)$ are the indices of the nearest 4 pixels to
$\theta$:
%%%%%%%%%%%%%%%%%%%%%%%%%%%%%%%%%%%%
\begin{equation}
(\Omega_{i_1},\Omega_{i_2},\Omega_{i_3},\Omega_{i_4}) = \cases{
(\Omega_0,\Omega_1,\Omega_2,\Omega_3) & $(0<\theta_x<\frac{\theta_{\rm pix}}{2}, 0<\theta_y<\frac{\theta_{\rm pix}}{2})$ \cr
(\Omega_0,\Omega_5,\Omega_4,\Omega_3)& $(0<\theta_x<\frac{\theta_{\rm pix}}{2}, -\frac{\theta_{\rm pix}}{2} < \theta_y <0)$ \cr
(\Omega_0,\Omega_5,\Omega_6,\Omega_7)& $(-\frac{\theta_{\rm pix}}{2} < \theta_x <0, -\frac{\theta_{\rm pix}}{2} < \theta_y <0)$ \cr
(\Omega_0,\Omega_1,\Omega_8,\Omega_7)& $(-\frac{\theta_{\rm pix}}{2} < \theta_x <0, 0<\theta_y<\frac{\theta_{\rm pix}}{2})$ \cr
}.	
\end{equation}
%%%%%%%%%%%%%%%%%%%%%%%%%%%%%%%%%%%%

Since the resulting effective PSF also depends on $\theta$ and
$\theta_{\rm g}$, we compute the PSF width appropriately averaged over
$\theta$ and $\theta_{\rm g}$ in the following.  In our analytic model
(\S \ref{sec:analytic}), we compute the expected $\Omega'(A')$ and
$N'_{\rm gal}(A')$ under the presence of the FIR contamination of
galaxies. We note that the effective PSF widths are slightly different
for $\Omega'(A')$ and $N'_{\rm gal}(A')$. This is because the extinction
contaminated by the FIR intensities, $A'$, is always evaluated at the
position of the galaxies, {\it i.e.,} $\theta=\theta_{\rm g}$, for
$N'_{\rm gal}(A')$, while this is not the case for $\Omega'(A')$.
Therefore we separately derive the effective PSF widths for
$\Omega'(A')$ and $N'_{\rm gal}(A')$.  We denote these effective PSF
widths as $\sigma_{{\rm eff},\Omega}$ and $\sigma_{{\rm eff},N}$.
 
Now let us calculate $\sigma_{{\rm eff},\Omega}$. Since $\theta$ and
$\theta_{\rm g}$ are independent for computing $\Omega'(A')$, we
calculate the intensity of galaxies averaged over $\theta$ and
$\theta_{\rm g}$ as
%%%%%%%%%%%%%%%%%%%%%%%%%%%%%%%%
\begin{equation}
 \bar{I} = \frac{1}{\Omega_{\rm pix}^2}
 \int_{\Omega_0}d\mathbf{\theta} \int_{\Omega_0}d\mathbf{\theta}_{\rm g}
I_{\rm{CIC}}(\mathbf{\theta},\mathbf{\theta}_{\rm g}).
\label{eq:ICIC-Omega}
\end{equation}
%%%%%%%%%%%%%%%%%%%%%%%%%%%%%%%%
We define $\sigma_{{\rm eff},\Omega}$ as 
%%%%%%%%%%%%%%%%%%%%%%%%%%%%%%%%
\begin{equation}
\frac{f}{2\pi \sigma^2_{\rm{eff},\Omega}} \equiv \bar{I},
\end{equation}
%%%%%%%%%%%%%%%%%%%%%%%%%%%%%%%%
and this leads to
%%%%%%%%%%%%%%%%%%%%%%%%%%%%%%%%
\begin{equation}
\sigma_{{\rm eff},\Omega} = \frac{4}{\sqrt{\pi}} \frac{\Omega_{\rm pix}}{\sigma_{\rm int}}
\frac{1}{6F(s)-5F(0)-2F(-s)+F(-2s)},
\label{eq:sigma-eff-Omega}
\end{equation}
%%%%%%%%%%%%%%%%%%%%%%%%%%%%%%%%
where
%%%%%%%%%%%%%%%%%%%%%%%%%%%%%%%%
\begin{equation}
F(x) = \int {\rm erf}(x)dx = x~{\rm erf}(x)+\frac{e^{-x^2}}{\sqrt{\pi}},
\end{equation}
%%%%%%%%%%%%%%%%%%%%%%%%%%%%%%%%
$s=\theta_{\rm pix}/\sqrt{2}\sigma_{\rm int}$, and ${\rm erf}(x)$
denotes the error function.

Similarly, considering that $\theta=\theta_{\rm g}$, we define
$\sigma_{{\rm eff},N}$ as
%%%%%%%%%%%%%%%%%%%%%%%%%%%%%%%%
\begin{equation}
\frac{f}{2\pi \sigma^2_{{\rm eff},N}} \equiv \frac{1}{\Omega_{\rm pix}}
 \int_{\Omega_0}I_{\rm{CIC}}(\mathbf{\theta}_{\rm g},\mathbf{\theta}_{\rm g}) d\mathbf{\theta}_{\rm g}.
 \label{eq:ICIC-N}
\end{equation}
%%%%%%%%%%%%%%%%%%%%%%%%%%%%%%%%
Equation (\ref{eq:ICIC-N}) is reduced to
%%%%%%%%%%%%%%%%%%%%%%%%%%%%%%%%
\begin{equation}
\sigma_{{\rm eff}, N} = \frac{\Omega_{\rm pix}}{\sqrt{8\pi \mathcal{R}}},
\label{eq:sigma-eff-N}
\end{equation}
%%%%%%%%%%%%%%%%%%%%%%%%%%%%%%%%
where
%%%%%%%%%%%%%%%%%%%%%%%%%%%%%%%%
\begin{equation}
\frac{2\mathcal{R}}{\sigma^2_{\rm int}} = 
\left[ J_1\left(-\frac{\theta_{\rm pix}}{2}\right) - J_2\left(-\frac{\theta_{\rm pix}}{2}\right)\right]^2 
+ 2J_1\left(-\frac{\theta_{\rm pix}}{2}\right)J_2\left(\frac{\theta_{\rm pix}}{2}\right)-
2J_2\left(-\frac{\theta_{\rm pix}}{2}\right) J_2\left(\frac{\theta_{\rm pix}}{2}\right)+
J_2\left(\frac{\theta_{\rm pix}}{2}\right)J_2\left(\frac{\theta_{\rm pix}}{2}\right),
\end{equation}
%%%%%%%%%%%%%%%%%%%%%%%%%%%%%%%%
%%%%%%%%%%%%%%%%%%%%%%%%%%%%%%%%
\begin{eqnarray}
J_1(x) &=& \left[ F(b+s)-F\left(b+\frac{s}{2}\right)-F(b)+F\left(b-\frac{s}{2}\right)\right], \\
J_2(x) &=& \frac{1}{s} \left[ G(b+s) - G\left(b+\frac{s}{2}\right) -G(b) + G\left(b-\frac{s}{2}\right)\right]
-\frac{1}{2} \left[ F\left(b+\frac{s}{2}\right) - F\left(b-\frac{s}{2}\right) \right] \\
&~& +  \frac{1}{2s} \left[ {\rm erf}(b+s) -{\rm erf}\left(b+\frac{s}{2}\right)
-{\rm erf} (b)+ {\rm erf} \left(b-\frac{s}{2}\right)\right], \\
G(x) &=& \int x~{\rm erf}(x)dx = \frac{1}{2}\left[ x^2{\rm erf}(x) +
 \frac{1}{\sqrt{\pi}} xe^{-x^2} -\frac{1}{2}{\rm erf}(x)\right],
\end{eqnarray}
%%%%%%%%%%%%%%%%%%%%%%%%%%%%%%%%
and $b \equiv x/\sqrt{2}\sigma_{\rm int}$.

The right panel of Figure \ref{fig:sigma-eff} shows the equations
(\ref{eq:sigma-eff-Omega}) and (\ref{eq:sigma-eff-N}) as functions of
$\sigma_{\rm int}$, which are adopted to equation (\ref{eq:P_1}) in the
analytic model presented in Appendix \ref{app:analytic-detail}.  In
numerical simulations in \S \ref{sec:simulation}, we adopted
$\sigma_{\rm int}=2'.21$, which reproduces the effective resolutions
$\sigma_{{\rm eff},\Omega}$ and $\sigma_{{\rm eff}, N}$ both similar to
the SFD angular resolution ${\rm FWHM} = 6'.1$.
%%%%%%%%%%%%%%%%%%%%%%%%%%%%%%%%%%%%%%%%%%%%%%%%%%%
%%%%%%%%%%%%%%%%%%%%%%%%%%%%%%%%%%%%%%%%%%%%%%%%%%%%

%%%%%%%%%%%%%%%%%%%%%%%%%%%%%%%%%%%%%%%%

%%%%%%%%%%%%%%%%%%%%%%%%%%%%%%%%%%%%%%%%
\section{Analytic model neglecting spatial clustering of galaxies}
\label{app:analytic-detail}
Assume that galaxies are randomly distributed over the pixel, and denote
the expected number of the galaxies of the {\it true} (albeit
unobservable) apparent magnitude $m_{\rm true}$ being $m_{\rm
min}<m_{\rm true}<m_{\rm max}$ by $\overline{N}$. Then the probability
that the pixel has $N$ galaxies obeys the Poisson distribution:
%%%%%%%%%%%%%%%%%%%%%%%%%%%%%%%%%%%%%%%%
\begin{equation}
\label{eq:Poisson}
P_{\rm{Poisson}}(N|\overline{N}) 
=  \frac{\overline{N}^N \exp(-\overline{N})}{N!}.
\end{equation}
%%%%%%%%%%%%%%%%%%%%%%%%%%%%%%%%%%%%%%%%
Here we assume that the area of all the pixels of the dust map is equal.
Then the joint probability is the product of the conditional probability
that the total FIR contamination in the pixel is $\Delta A$, given that
there are $N$ galaxies and that the probability that the pixel has $N$
galaxies:
%%%%%%%%%%%%%%%%%%%%%%%%%%%%%%%%%%%%%%%%
\begin{equation}
\label{eq:P_joint}
P_{\rm{joint}}(\Delta A,N) 
= P_N(\Delta A)P_{\rm{Poisson}}(N|\overline{N}) .
\end{equation}
%%%%%%%%%%%%%%%%%%%%%%%%%%%%%%%%%%%%%%%%

The conditional probability $P_N(\Delta A)$ can be computed recursively.
When there is no galaxy in a pixel ($N=0$), $\Delta A$ should vanish:
%%%%%%%%%%%%%%%%%%%%%%%%%%%%%%%%%%%%%%%%
\begin{equation}
P_0(\Delta A) = \delta_D ( \Delta A),
\label{eq:P_0}
\end{equation}
%%%%%%%%%%%%%%%%%%%%%%%%%%%%%%%%%%%%%%%%
where $\delta_D$ is the 1-dimensional Dirac delta function.  We compute
$P_1(\Delta A)$ from the differential number count of galaxy magnitude
and the PDF of the FIR to $r$-band flux ratio as discussed later in
detail.  Then $P_N(\Delta A)$ for $N \geq 2$ should satisfy the
following recursive equation:
%%%%%%%%%%%%%%%%%%%%%%%%%%%%%%%%%%%%%%%%
\begin{equation}
\label{eq:P_N}
P_N(\Delta A) = \int_0^{\Delta A} dx P_1(x) P_{N-1}(\Delta A-x).
\end{equation}
%%%%%%%%%%%%%%%%%%%%%%%%%%%%%%%%%%%%%%%%

Finally the PDF of the total contamination in a
pixel, $P(\Delta A)$, is given as
%%%%%%%%%%%%%%%%%%%%%%%%%%%%%%%%%%%%%%%%
\begin{eqnarray}
\label{eq:P_pixel}
P(\Delta A) &=& \sum_{N=0}^{\infty} P_{\rm{joint}}(\Delta A,N) 
= \sum_{N=0}^{\infty}P_N(\Delta A)P_{\rm{Poisson}}(N|\overline{N}) .
\end{eqnarray}
%%%%%%%%%%%%%%%%%%%%%%%%%%%%%%%%%%%%%%%%
Note therefore that $P_{\rm{joint}}(\Delta A,N)$ and $P(\Delta A)$ are
computed in a straightforward fashion once the two inputs, $P_1(\Delta
A)$ and $\overline{N}$, are specified from the observed data.

Next let us proceed to compute $\Omega^{\prime}(A^{\prime})$ and
$N^{\prime}(A^{\prime})$ according to this model.  Since SFD subtracted
the mean FIR contamination in a pixel in constructing the map, we also
subtract its theoretical counterpart:
%%%%%%%%%%%%%%%%%%%%%%%%%%%%%%%%%%%%%%%%
\begin{equation}
\label{eq:average-DeltaA}
\overline{\Delta A} = \int_0^{\infty}  
d(\Delta A) \Delta A P(\Delta A), 
\end{equation}
%%%%%%%%%%%%%%%%%%%%%%%%%%%%%%%%%%%%%%%%
from the FIR contamination $\Delta A$ in each pixel. So the extinction
contaminated by the galaxy emission is now given by
%%%%%%%%%%%%%%%%%%%%%%%%%%%%%%%%%%%%%%%%
\begin{equation}
A^{\prime} = A+\Delta A-\overline{\Delta A} .
\end{equation}
%%%%%%%%%%%%%%%%%%%%%%%%%%%%%%%%%%%%%%%%
Therefore, the probability that a pixel with the {\em true} extinction
$A$ is observed as $A^{\prime}$ due to the FIR contamination is given by
$P(\Delta A)=P(A^{\prime}-A+\overline{\Delta A})$.  Finally we obtain
the expected observed distribution function of sky area,
$\Omega^{\prime}(A^{\prime})$ as
%%%%%%%%%%%%%%%%%%%%%%%%%%%%%%%%%%%%%%%%
\begin{eqnarray}
\label{eq:Omega^prime}
\Omega^{\prime}(A^{\prime})
&=&
\int_0^{\infty} dA \int_0^{\infty} d(\Delta A) \Omega(A) P(\Delta A) 
\delta_D\big(A^{\prime}-(A+\Delta A-\overline{\Delta
A})\big) 
=
\int_0^{A^{\prime}+\overline{\Delta A}} dA 
\Omega(A) P(A^{\prime}-A+\overline{\Delta A}).
\end{eqnarray}
%%%%%%%%%%%%%%%%%%%%%%%%%%%%%%%%%%%%%%%%
We can similarly derive the expression for
$N_{\rm{gal}}^{\prime}(A^{\prime})$, the number distribution of the
galaxies located in the pixels of the extinction $A^\prime$, as follows.

Since we assume that the area of each pixel is the same and equal to
$\Omega_{\rm{pixel}}$, the number of pixels that have the {\em true}
extinction in the range of $A$ and $A+dA$ is
%%%%%%%%%%%%%%%%%%%%%%%%%%%%%%%%%%%%%%%
\begin{equation}
\label{eq:pixel-number}
N_{\rm{pixel}}(A)dA = \frac{\Omega(A)dA}{\Omega_{\rm{pixel}}}.
\end{equation}
%%%%%%%%%%%%%%%%%%%%%%%%%%%%%%%%%%%%%%%
Thus the expected number distribution of galaxies in a pixel that
suffers from the FIR contamination of $\Delta A$ is
%%%%%%%%%%%%%%%%%%%%%%%%%%%%%%%%%%%%%%%
\begin{equation}
\overline{N}(\Delta A) 
= \sum_{N=0}^{\infty} N P_{\rm{joint}}(\Delta A,N).
\end{equation}
%%%%%%%%%%%%%%%%%%%%%%%%%%%%%%%%%%%%%%%
Therefore, the number distribution of  galaxies,
$N_{\rm{gal}}^{\prime}(A^{\prime})$, is given as
%%%%%%%%%%%%%%%%%%%%%%%%%%%%%%%%%%%%%%%
\begin{eqnarray}
\label{eq:N^prime}
N_{\rm{gal}}^{\prime}(A^{\prime})
&=&
\int_0^{\infty} dA \int_0^{\infty} d(\Delta A) N_{\rm{pixel}}(A) 
\overline{N}(\Delta A) 
 \delta_D\big(A^{\prime}-(A+\Delta A-\overline{\Delta A})\big) \cr
&=& \int_0^{A^\prime +\overline{\Delta A}}d(\Delta A) 
N_{\rm{pixel}}(A^{\prime}-\Delta A+\overline{\Delta A})
\overline{N}(\Delta A) .
\end{eqnarray}
%%%%%%%%%%%%%%%%%%%%%%%%%%%%%%%%%%%%%%%

While the above expression is correct for those galaxies with $m_{\rm
min}<m_{\rm true}<m_{\rm max}$, we cannot measure their true magnitude
$m_{\rm true}$ in reality, and one has to take into account the
selection effect carefully.  Consider a galaxy of $m_{\rm true}$ is
located in a pixel of the contaminated extinction of $A^{\prime}$. Then
its observed (uncorrected) magnitude is
%%%%%%%%%%%%%%%%%%%%%%%%%%%%%%%%%%%%%%%
\begin{equation}
m_{\rm uncorr}(A^\prime) = m_{\rm true} +A ,
\end{equation}
%%%%%%%%%%%%%%%%%%%%%%%%%%%%%%%%%%%%%%%
because its magnitude suffers from the true Galactic extinction $A$
alone, instead of $A^\prime$. This yields the {\it corrected} magnitude
relying on the contaminated extinction $A^{\prime}$:
%%%%%%%%%%%%%%%%%%%%%%%%%%%%%%%%%%%%%%%
\begin{eqnarray}
m_{\rm corr}(A^\prime) &=&
m_{\rm uncorr}(A^\prime) -A^\prime = m_{\rm true} +A-A^\prime
= m_{\rm true} - (\Delta A- \overline{\Delta A})
\end{eqnarray}
%%%%%%%%%%%%%%%%%%%%%%%%%%%%%%%%%%%%%%%
leading to the {\it over-correction} by the amount of
$\Delta A- \overline{\Delta A}$.

Therefore, those galaxies with $m_{\rm min}<m_{\rm corr}(A^\prime)<
m_{\rm max}$ indeed correspond to
%%%%%%%%%%%%%%%%%%%%%%%%%%%%%%%%%%%%%%%
\begin{equation}
m_{\rm min} + (\Delta A- \overline{\Delta A})
 < m_{\rm true} < m_{\rm max}+(\Delta A- \overline{\Delta A}).
\end{equation}
%%%%%%%%%%%%%%%%%%%%%%%%%%%%%%%%%%%%%%%
In other words, the selection incorrectly excludes galaxies with $m_{\rm
min} <m_{\rm true} < m_{\rm min}+\Delta A-\overline{\Delta A}$, and
includes those with $m_{\rm max} <m_{\rm true} < m_{\rm max}+\Delta
A-\overline{\Delta A}$ because of the contamination of FIR galaxy
emission.

Given their differential number count with respect to magnitude, the
number of such galaxies can be computed as
%%%%%%%%%%%%%%%%%%%%%%%%%%%%%%%%%%%%%%%
\begin{eqnarray}
N_{\rm{ex,corr}}(\Delta A) 
&=& \int_{m_{\rm{min}}}^{m_{\rm{min}}+\Delta A-\overline{\Delta A}}
\frac{dn(<m)}{dm}dm, \\
N_{\rm{in,corr}}(\Delta A) 
&=& \int_{m_{\rm{max}}}^{m_{\rm{max}}+\Delta A-\overline{\Delta A}}
\frac{d\overline{n}(<m)}{dm}dm. 
\label{eq:cr_exc_inc}
\end{eqnarray}
%%%%%%%%%%%%%%%%%%%%%%%%%%%%%%%%%%%%%%%
We adopt a power-law fit with a slope $\gamma$ (see
Fig. \ref{fig:magnitude-distribution}) for the differential number
counts of galaxies in a pixel that contains $N$ and $\overline{N}$
galaxies:
%%%%%%%%%%%%%%%%%%%%%%%%%%%%%%%%%%%%%%%
\begin{eqnarray}
\frac{dn(<m)}{dm} &=& \frac{N\gamma 10^{\gamma m}\ln 10}
{10^{\gamma m_{\rm{max}}}-10^{\gamma m_{\rm{min}}}}, \\
\frac{d\overline{n}(<m)}{dm} 
&=& \frac{\overline{N}\gamma 10^{\gamma m}\ln 10}
{10^{\gamma m_{\rm{max}}}-10^{\gamma m_{\rm{min}}}} .
\label{eq:cr_dndm}
\end{eqnarray}
%%%%%%%%%%%%%%%%%%%%%%%%%%%%%%%%%%%%%%%
The excluded number should be normalized for the actual number of
galaxies, $N$, instead of $\overline{N}$, in the pixel. Nevertheless 
the included number is not correlated to $N$ in the Poisson distributed
assumption, and thus should be normalized for $\overline{N}$.

Therefore we obtain finally the number distribution of galaxies after
correcting for the contaminated extinction $A^{\prime}$ as
%%%%%%%%%%%%%%%%%%%%%%%%%%%%%%%%%%%%%%%
\begin{eqnarray}
N_{\rm{gal,corr}}^{\prime}(A^{\prime})
&=&
\int_0^{\infty} dA \int_0^{\infty} d(\Delta A) N_{\rm{pixel}}(A)
 [\overline{N}(\Delta A)-N_{\rm{ex,corr}}(\Delta A)
+N_{\rm{in,corr}}(\Delta A)] 
 \delta_D\big(A^{\prime}-(A+\Delta A-\overline{\Delta A})\big)
 \cr
&=& \int_0^{A^\prime +\overline{\Delta A}} d(\Delta A) 
N_{\rm{pixel}}(A^{\prime}-\Delta A+\overline{\Delta A}) 
\times [\overline{N}(\Delta A)-N_{\rm{ex,corr}}(\Delta A)
+N_{\rm{in,corr}}(\Delta A)] .
\label{eq:number-after}
\end{eqnarray}
%%%%%%%%%%%%%%%%%%%%%%%%%%%%%%%%%%%%%%%

Similarly, the number distribution of galaxies {\it before} correcting for 
the contaminated extinction $A^{\prime}$, {\it i.e.,} with
$m_{\rm min}-A <m_{\rm true} < m_{\rm max}-A$, is given as
%%%%%%%%%%%%%%%%%%%%%%%%%%%%%%%%%%%%%%%
\begin{eqnarray}
N_{\rm{gal,uncorr}}^{\prime}(A^{\prime})
&=&
\int_0^{\infty} dA \int_0^{\infty} d(\Delta A) N_{\rm{pixel}}(A) 
[\overline{N}(\Delta A)-N_{\rm{ex,uncorr}}(A)+N_{\rm{in,uncorr}}(A)] \cr
&\times& \delta_D\big(A^{\prime}-(A+\Delta A-\overline{\Delta A})\big),
\label{eq:number-before}
\end{eqnarray}
%%%%%%%%%%%%%%%%%%%%%%%%%%%%%%%%%%%%%%%
where
%%%%%%%%%%%%%%%%%%%%%%%%%%%%%%%%%%%%%%%
\begin{equation}
N_{\rm{ex,uncorr}}(A) 
=\int_{m_{\rm{max}}-A}^{m_{\rm{max}}}\frac{dn(<m)}{dm}dm,
\label{eq:un_exc}
\end{equation}
%%%%%%%%%%%%%%%%%%%%%%%%%%%%%%%%%%%%%%%
and
%%%%%%%%%%%%%%%%%%%%%%%%%%%%%%%%%%%%%%%
\begin{equation}
N_{\rm{in,uncorr}}(A) 
= \int_{m_{\rm{min}}-A}^{m_{\rm{min}}}\frac{d\overline{n}(<m)}{dm}dm .
\label{eq:un_inc}
\end{equation}
%%%%%%%%%%%%%%%%%%%%%%%%%%%%%%%%%%%%%%%

In order to proceed further, we need an expression for the PDF of the
FIR contamination due to a single galaxy, $P_1(\Delta A)$.  The mock
simulations presented in \S \ref{sec:simulation} convert the $r$-band
magnitude, $m_r$, of each mock galaxy into its 100 ${\mum}$ flux from
the FIR/optical luminosity ratio $y$ as
%%%%%%%%%%%%%%%%%%%%%%%%%%%%%%%%%%%%%%%%
\begin{equation}
f_{100\mu\mathrm{m}}(m_r,y) = y f_0 10^{-0.4m_r},
\label{eq:I(m,y)}
\end{equation}
%%%%%%%%%%%%%%%%%%%%%%%%%%%%%%%%%%%%%%%%
where $f_0=3631{\rm Jy}$, and $y$ is assumed to obey the log-normal PDF
$P_{\rm{ratio}}$ given by equation (\ref{eq:log-normal}).  In the
present analytic model, we further assume that the differential number
count of galaxies in $r$-band obeys
%%%%%%%%%%%%%%%%%%%%%%%%%%%%%%%%%%%%%%%%
\begin{equation}
P_{\rm{mag}}(m_r) =\frac{\gamma_r 10^{\gamma_r m_r}\ln 10}
{10^{\gamma_r m_{r,\rm{max}}}- 10^{\gamma_r m_{r,\rm{min}}}},
\label{eq:P(m)}
\end{equation}
%%%%%%%%%%%%%%%%%%%%%%%%%%%%%%%%%%%%%%%%
where $m_{r,\rm{max}}$ and $m_{r,\rm{min}}$ denote the upper and lower
limits of the magnitude, and $\gamma_r$ is the power-law index.

Once $P_{\rm{mag}}(m_r)$ and $P_{\rm{ratio}}(y)$ are given, 
the PDF of 100$\mum$ flux from a single galaxy is computed as
%%%%%%%%%%%%%%%%%%%%%%%%%%%%%%%%%%%%%%%%
\begin{eqnarray}
P_{\rm flux}(f) 
&=& \int dy \int dm_r ~ P_{\rm{mag}}(m_r) P_{\rm{ratio}}(y) 
\delta_D \big(f-f_{100{\mum}}(m_r,y)\big).
\label{eq:P_flux}
\end{eqnarray}
%%%%%%%%%%%%%%%%%%%%%%%%%%%%%%%%%%%%%%%%
With the PDFs of equations (\ref{eq:P(m)}) and (\ref{eq:log-normal}),
 $P_{\rm flux}(f)$ reduces to
%%%%%%%%%%%%%%%%%%%%%%%%%%%%%%%%%%%%%%%%
\begin{eqnarray}
P_{\rm flux}(f) &=& 
K\left(\frac{f}{f_0}\right)^{-1-\frac{5}{2}\gamma_r}
\left[ \mathrm{erf}\big(s_{\rm{max}}(f)\big)
-\mathrm{erf}\big(s_{\rm{min}}(f)\big)\right],
\label{eq:P_flux-2}
\end{eqnarray}
%%%%%%%%%%%%%%%%%%%%%%%%%%%%%%%%%%%%%%%%
where $\mathrm{erf}(x)$ denotes the error function, and $K$,
$s_{\rm{max}}$ and $s_{\rm{min}}$ are defined as
%%%%%%%%%%%%%%%%%%%%%%%%%%%%%%%%%%%%%%%%
\begin{eqnarray}
K 
&\equiv&
\frac{5\gamma_r 10^{\frac{5}{2}\mu \gamma_r}}
{4f_0(10^{\gamma_r m_{r,\rm{max}}}-10^{\gamma_r m_{r,\rm{min}}})} 
\exp \left[ \frac{25}{8}\sigma^2 \gamma_r^2 (\ln 10)^2 \right],
 \\
s_{\rm{max}}(f) 
&\equiv&
\frac{1}{\sqrt{2\sigma^2}} \bigg[ 0.4m_{r,\rm{max}}- \mu 
+\log_{10}\left(\frac{f}{f_0}\right)- \frac{5}{2}\sigma^2 \gamma_r \ln 10\bigg], \\
s_{\rm{min}}(f)
& \equiv &
\frac{1}{\sqrt{2\sigma^2}} \bigg[ 0.4m_{r,\rm{min}}- \mu 
 + \log_{10}\left(\frac{f}{f_0}\right) 
- \frac{5}{2}\sigma^2 \gamma_r \ln 10\bigg].
\end{eqnarray}
%%%%%%%%%%%%%%%%%%%%%%%%%%%%%%%%%%%%%%%%
Incidentally, $P_{\rm flux}(f)$ turns out to be well approximated by a
log-normal function also, but we use equation (\ref{eq:P_flux-2}) to be
precise.  Considering that the mock galaxies with flux larger than $f_{\rm lim}$ are removed 
and do not contaminate, $P_1(\Delta A)$ is calculated as
%%%%%%%%%%%%%%%%%%%%%%%%%%%%%%%%%%%%%%%%
\begin{eqnarray}
P_1(\Delta A) = \delta_D(\Delta A)\int_{f_{\rm lim}}^{\infty} P_{\rm flux}(f)df +
 \frac{1}{C}\Theta(Cf_{\rm lim}-\Delta A) P_{\rm flux}\left(\frac{\Delta A}{C}\right),
\label{eq:P_1}
\end{eqnarray}
%%%%%%%%%%%%%%%%%%%%%%%%%%%%%%%%%%%%%%%%
where $C \equiv k_r p /\Omega_{\rm{pix,eff}}$ is a conversion factor
from the FIR flux to the $r$-band extinction.  We adopt
$\Omega_{\rm{pix,eff}}=2\pi \sigma_{\rm{eff}}^2$ as the effective area
of a pixel, where $\sigma_{\rm{eff}}$ is the Gaussian width
corresponding to the effective angular resolution, which is given in
Appendix \ref{app:psf}. We adopt equation (\ref{eq:sigma-eff-Omega}) for
calculating $\Omega'(A')$, and (\ref{eq:sigma-eff-N}) for $N'(A')$.

An analytic model that we present in this paper neglects the spatial
clustering of galaxies, but it is, at least partially, incorporated by
the assigned value of 100{\mum} flux for each $r$-band selected
galaxy. The interpretation is slightly subtle, but we would like to
emphasize that the neglect of the spatial clustering in our analytic
model is not serious in practice as discussed in \S
\ref{sec:discussion}.
\bigskip

%%%%%%%%%%%%%%%%%%%%%%%%%%%%%%%%%%%%%%%%%%%%%%
%    References 
%%%%%%%%%%%%%%%%%%%%%%%%%%%%%%%%%%%%%%%%%%%%%%
\bibliographystyle{apj}
\bibliography{modeling-SFDanomaly.bbl}

\end{document}